\documentclass[a4paper]{article}

\usepackage{a4wide}
\usepackage[UKenglish]{babel}
\usepackage{authblk}
\usepackage{graphicx,subfigure,caption}
\graphicspath{{./}{figures/}}
\usepackage[colorlinks=true,linkcolor=blue,citecolor=red]{hyperref}
\usepackage{xcolor}
\usepackage{amssymb,amsthm,empheq,bbold}
\usepackage[inline]{enumitem}
\usepackage[ruled,vlined,linesnumbered]{algorithm2e}

\newcommand{\R}{\mathbb{R}}
\newcommand{\Real}[1]{\Re\text{e}\,#1}

\newtheorem{assumption}{Assumption}[section]
\theoremstyle{remark}\newtheorem{remark}[assumption]{Remark}

\allowdisplaybreaks

\title{A statistical mechanics approach to macroscopic limits of car-following traffic dynamics}
\author[1]{Felisia Angela Chiarello}
\author[2]{Benedetto Piccoli}
\author[1]{Andrea Tosin}
\affil[1]{{\footnotesize Department of Mathematical Sciences ``G. L. Lagrange'', Politecnico di Torino, Torino, Italy}}
\affil[2]{{\footnotesize Department of Mathematical Sciences, Rutgers University, Camden NJ, USA}}

\date{}

\begin{document}
\maketitle
	
\begin{abstract}
We study the derivation of macroscopic traffic models from car-following vehicle dynamics by means of hydrodynamic limits of an Enskog-type kinetic description. We consider the superposition of Follow-the-Leader (FTL) interactions and relaxation towards a traffic-dependent Optimal Velocity (OV) and we show that the resulting macroscopic models depend on the relative frequency between these two microscopic processes. If FTL interactions dominate then one gets an inhomogeneous Aw-Rascle-Zhang model, whose (pseudo) pressure and stability of the uniform flow are precisely defined by some features of the microscopic FTL and OV dynamics. Conversely, if the rate of OV relaxation is comparable to that of FTL interactions then one gets a Lighthill-Whitham-Richards model ruled only by the OV function. We further confirm these findings by means of numerical simulations of the particle system and the macroscopic models. Unlike other formally analogous results, our approach builds the macroscopic models as physical limits of particle dynamics rather than assessing the convergence of microscopic to macroscopic solutions under suitable numerical discretisations.

\medskip


\medskip

\end{abstract}
	
\section{Introduction}
A very popular microscopic description of vehicular traffic is based on Follow-the-Leader (FTL) dynamics~\cite{chandler1958OR,gazis1961OR}, which assume that the acceleration of a vehicle is proportional to the relative speed with the leading vehicle and inversely proportional to the relative distance:
\begin{equation}
	\begin{cases}
		\dot{x}_j=v_j \\
		\dot{v}_j=\lambda_0\dfrac{v_{j+1}-v_j}{(x_{j+1}-x_j)^{1+\gamma}},
	\end{cases}
	\qquad
	j=1,\,2,\,\dots,
	\label{eq:FTL}
\end{equation}
where $x_j=x_j(t)$, $v_j=v_j(t)$ are the position and speed, respectively, of the $j$th vehicle at time $t$, $j+1$ is the vehicle leading vehicle $j$ and $\lambda_0>0$, $\gamma\geq 0$ are parameters. This model expresses the tendency of a vehicle to relax its speed towards that of the leading vehicle more or less promptly depending on the free space ahead (headway).

A different line of microscopic models is instead based on Optimal Velocity (OV) dynamics~\cite{bando1995PRE}, which assume that vehicles aim to align their speed to a headway-dependent optimal speed:
\begin{equation}
	\begin{cases}
		\dot{x}_j=v_j \\
		\dot{v}_j=a\bigl(V(x_{j+1}-x_j)-v_j\bigr),
	\end{cases}
	\qquad
	j=1,\,2,\,\dots,
	\label{eq:OV}
\end{equation}
$V$ being the prescribed optimal speed function and $a>0$ a relaxation parameter.

Putting FTL and OV dynamics together one obtains a general car-following model~\cite{aw2002SIAP}:
\begin{equation}
	\begin{cases}
		\dot{x}_j=v_j \\
		\dot{v}_j=\lambda_0\dfrac{v_{j+1}-v_j}{(x_{j+1}-x_j)^{1+\gamma}}+a\bigl(V(x_{j+1}-x_j)-v_j\bigr),
	\end{cases}
	\qquad
	j=1,\,2,\,\dots,
	\label{eq:FTL-OV}
\end{equation}
in which two competing trends, both mediated by the headway, drive the relaxation of a vehicle's speed towards either the speed of the leading vehicle or the optimal speed.

Great efforts have been devoted in the literature to establish the macroscopic counterparts of these models. In~\cite{aw2002SIAP} the authors show that model~\eqref{eq:FTL-OV} converges to the inhomogeneous Aw-Rascle-Zhang (ARZ) model~\cite{greenberg2001SIAP,rascle2002MCM}:
\begin{equation}
	\begin{cases}
		\partial_t\rho+\partial_x(\rho u)=0 \\
		\partial_t(u+p(\rho))+u\partial_x(u+p(\rho))=a(V(\rho)-u),
	\end{cases}
	\label{eq:inhom_ARZ.original}
\end{equation}
were $\rho$, $u$ are the traffic density and mean speed, respectively, and $p(\rho)$ is a (pseudo) traffic pressure sometimes also called hesitation function~\cite{ramadan2021SEMAI-SIMAI}. The principle used in~\cite{aw2002SIAP} consists in interpreting a discrete-in-time version of~\eqref{eq:FTL-OV} as a space-time discretisation of~\eqref{eq:inhom_ARZ.original} for a suitably chosen traffic pressure $p$ and in showing that the discrete solutions thereby produced converge to weak solutions of~\eqref{eq:inhom_ARZ.original} as the discretisation parameters become infinitesimal. Along a similar line of thought, in~\cite{difrancesco2017MBE} the authors prove that the original (homogeneous) ARZ model~\cite{aw2000SIAP,zhang2002TRB}, i.e. model~\eqref{eq:inhom_ARZ.original} with $a=0$, can be viewed as the limit of an FTL model of type~\eqref{eq:FTL} for particular choices of $\gamma$ and non-constant $\lambda_0$ suitably related to the traffic pressure $p$. In~\cite{colombo2014RSMUP,difrancesco2015ARMA} it is also shown that the simpler first-order FTL model
\begin{equation}
	\dot{x}_j=V\!\left(\frac{\ell}{x_{j+1}-x_j}\right), \qquad j=1,\,2,\,\dots,
	\label{eq:first_order.FTL}
\end{equation}
where $\ell>0$ is a vehicle's length, converges to the Lighthill-Whitham-Richards (LWR) macroscopic traffic model~\cite{lighthill1955PRSL,richards1956OR}
\begin{equation}
	\partial_t\rho+\partial_x(\rho V(\rho))=0
	\label{eq:LWR.original}
\end{equation}
in the sense outlined above when $\ell\to 0^+$. These types of limits have been applied also to justify macroscopic traffic models with non-local flux~\cite{chiarello2020SIAP,goatin2017CMS} or macroscopic traffic models on networks~\cite{cristiani2016NHM}.

The limit procedure underlying these results actually shows that the microscopic models are consistent \textit{discretisations} of the macroscopic ones. Not by chance, the obtained convergence of the discrete to the continuous solutions is often exploited as a tool to prove the existence of solutions to the macroscopic models and to build alternative schemes for their numerical approximation. Yet, the question of obtaining inhomogeneous ARZ or LWR models from first principles, i.e. not by comparing their solutions with those of selected microscopic models used as numerical discretisation (which is a sort of educated guess) but by recovering them genuinely as physical limits of certain particle dynamics, remains open.

In this paper, we take this second point of view and we show that the aforementioned macroscopic models emerge naturally as \textit{hydrodynamic limits} of combined FTL and OV particle dynamics under precise physical assumptions on their mutual relationships. In particular, the macroscopic description switches from the inhomogeneous ARZ model~\eqref{eq:inhom_ARZ.original} to the LWR model~\eqref{eq:LWR.original} in consequence of minimal qualitative changes of the \textit{same particle dynamics}. These changes involve the rate of OV relaxations compared to that of FTL interactions. Such a fundamental structural aspect remains hidden in the limit procedures based on the comparison of microscopic and macroscopic solutions, which suggest instead that e.g., a difference as relevant as the one between the microscopic dynamics~\eqref{eq:FTL-OV} and~\eqref{eq:first_order.FTL} is required to pass from~\eqref{eq:inhom_ARZ.original} to~\eqref{eq:LWR.original}. Another interesting aspect of our approach is that some hallmarks of the macroscopic models, such as e.g., the traffic pressure $p$ in~\eqref{eq:inhom_ARZ.original}, turn out to be naturally linked to general features of the particle dynamics. Also this aspect is sometimes hidden in the limit procedures discussed above, for the latter may require to formulate \textit{a priori} some specific parallelism between the corresponding terms of the microscopic and macroscopic models in order to obtain the convergence of either solution in the limit.

Our approach is based on a kinetic description of joint FTL and OV particle dynamics. The description of traffic flow by appealing to principles of statistical mechanics and kinetic theory dates back to the pioneering works~\cite{prigogine1960OR,prigogine1971BOOK} and in successive times has been the object of a renewed interest. Without intending to review all the pertinent literature, we refer the interested reader to e.g.,~\cite{dimarco2020JSP,herty2007NHM,illner2009QAM,klar1997JSP,klar2000SIAP} and references therein. In particular, in~\cite{klar1997JSP,klar2000SIAP} the authors were the first to show that the method of moment equations applied to an Enskog-type kinetic description of generic traffic dynamics may formally lead to the (homogeneous) ARZ model. Their approach will be very much inspiring for us but at the same time we will improve it in the spirit of~\cite{dimarco2020JSP}. In particular, we will focus on a rigorous upscaling of microscopic interaction rules specifically motivated by the superposition of FTL and OV dynamics, which will allow us to obtain explicit closure relationships at the kinetic and macroscopic levels and to avoid partly heuristic ones like in~\cite{klar1997JSP,klar2000SIAP}. This requires in particular to face the collision operators of the kinetic equations, which are usually a highly non-trivial part of the theory but that we are nevertheless able to successfully manage.

In more detail, the paper is organised in three main sections that follow this introduction. In Section~\ref{sect:slow_OV} we show that the inhomogeneous ARZ model~\eqref{eq:inhom_ARZ.original} emerges as the hydrodynamic limit of the particle regime of \textit{slow} OV dynamics, i.e. the one in which the OV relaxation takes place at a \textit{lower} rate than FTL interactions. In Section~\ref{sect:fast_OV} we show instead that the LWR model~\eqref{eq:LWR.original} emerges as the hydrodynamic limit of the particle regime of \textit{fast} OV dynamics, i.e. the one in which the OV relaxation takes place at the \textit{same} rate as (notice: not necessarily at a \textit{higher} rate than) FTL interactions. In Section~\ref{sect:numerics} we simulate the particle model in either regime and we compare it with the appropriate macroscopic model deduced from the previously developed theory. Finally, in Section~\ref{sec:conclusions} we outline some conclusions.

\section{Slow OV dynamics}
\label{sect:slow_OV}
Let us consider a sufficiently large ensemble of indistinguishable vehicles, each of which is identified by its dimensionless position $X_t\in\R$ and dimensionless speed $V_t\in [0,\,1]$ at time $t>0$. Inspired by the FTL and OV dynamics discussed in the Introduction, we assume the following discrete-in-time stochastic particle model:
\begin{equation}
	\begin{cases}
		X_{t+\Delta{t}}=X_t+\Delta{t}V_t \\
		V_{t+\Delta{t}}=V_t+\Theta\lambda(h(\rho))(V_t^\ast-V_t)+\Xi a(V(h(\rho))-V_t),
	\end{cases}
	\label{eq:particle_model}
\end{equation}
where $\Delta{t}>0$ is a (small) time gap.

Here, $\Theta,\,\Xi\in\{0,\,1\}$ are independent binary random variables, which describe whether during the time gap $\Delta{t}$ a randomly chosen vehicle updates its speed because of either an FTL interaction with another randomly chosen vehicle, whose speed is denoted by $V_t^\ast$, or an OV relaxation towards the optimal speed represented by the prescribed function $V$. In more detail, we let
\begin{equation}
	\Theta\sim\operatorname{Bernoulli}(\Delta{t}), \qquad \Xi\sim\operatorname{Bernoulli}(\epsilon\Delta{t}),
	\label{eq:Bernoulli.slow_OV}
\end{equation}
thereby assuming that the probability for either update to happen is proportional to $\Delta{t}$. The parameter $\epsilon>0$ is used to differentiate the frequency of OV speed updates from that of FTL updates. In particular, here we assume that OV updates happen at a much lower frequency than FTL speed updates, i.e.
$$ \epsilon\ll 1. $$
This may be motivated by the fact that FTL speed updates result from vehicle-to-vehicle interactions taking place continuously in the traffic stream while OV speed updates only require the self-adaptation of one vehicle's speed to the optimal speed. Notice that we need $\Delta{t}\leq\min\{1,\,\frac{1}{\epsilon}\}=1$ for consistency.

Furthermore:
\begin{enumerate*}[label=(\roman*)]
\item $\rho\geq 0$ denotes the \textit{traffic density} in a neighbourhood of the point $X_t$, hence~\eqref{eq:particle_model} is a \textit{non-local} particle model;
\item the function $h:\R_+\to\R_+$ models the \textit{headway} of the vehicle in $X_t$;
\item $\lambda=\lambda(h(\rho))$ is, according to the terminology introduced in~\cite{gazis1961OR}, the \textit{sensitivity} of the driver in $X_t$;
\item $a>0$ is a relaxation rate.
\end{enumerate*}

Concerning the functions $V$, $h$, we make the following assumptions:
\begin{assumption}
We assume that the optimal speed $V$ is a function of the headway $h\geq 0$ with the following characteristics:
\begin{enumerate}[label=(\roman*)]
\item it is bounded between $0$ and $1$:
$$ 0\leq V(h)\leq 1, \qquad \forall\,h\geq 0; $$
\item it is differentiable and monotonically increasing:
$$ V'(h)>0, \qquad \forall\,h\geq 0. $$
\end{enumerate}

Moreover, we assume that the headway $h$ is a function of the traffic density $\rho\geq 0$ with the following characteristics:
\begin{enumerate}[label=(\roman*)]
\item it is non-negative:
$$ h(\rho)\geq 0, \qquad \forall\,\rho\geq 0; $$
\item it is differentiable and monotonically decreasing:
$$ h'(\rho)<0, \qquad \forall\,\rho\geq 0. $$
\end{enumerate}
\label{ass:Vh}
\end{assumption}

A common definition in the traffic literature, see e.g.,~\cite{gazis1961OR}, is
$$ h(\rho)=\frac{c}{\rho}, $$
where $c>0$ is a proportionality constant. A slight modification, which produces a bounded headway function also for small $\rho$, is
\begin{equation}
	h(\rho)=\frac{c}{1+\rho}.
	\label{eq:h.bounded}
\end{equation}
Then, getting inspiration from the FTL-OV microscopic model~\eqref{eq:FTL-OV} we may let
$$ \lambda(h(\rho))=\frac{\lambda_0}{h^{1+\gamma}(\rho)}. $$
Alternatively, we obtain a sensitivity function \textit{a priori} bounded for large $\rho$ by considering the correction
\begin{equation}
	\lambda(h(\rho))=\frac{\lambda_0}{1+h^{1+\gamma}(\rho)}.
	\label{eq:lambda.bounded}
\end{equation}

Conversely, the prototype of optimal speed $V$ is the one provided in~\cite{bando1995PRE}:
\begin{equation}
	V(h)=\tanh{(\alpha h)},
	\label{eq:V.tanh}
\end{equation}
where $\alpha>0$ is a parameter.

\subsection{Enskog-type kinetic description}
\label{sect:kinetic}
To give an aggregate statistical description of the particle model~\eqref{eq:particle_model} along the lines of statistical mechanics we introduce a so-called \textit{observable quantity} $\phi$, namely any function $\phi=\phi(x,v):\R\times [0,\,1]\to\R$ which may be computed out of the microscopic state $(x,v)$ of the vehicles and we average the dynamics~\eqref{eq:particle_model}. Assuming technically that $\phi$ is compactly supported in the variable $x$, in the continuous-time limit $\Delta{t}\to 0^+$ we formally obtain by standard arguments, see e.g.,~\cite{pareschi2013BOOK}, the kinetic description
\begin{align}
	\begin{aligned}[b]
		\frac{d}{dt}\int_{\R}\int_0^1&\phi(x,v)f(x,v,t)\,dv\,dx+\int_{\R}\int_0^1\phi(x,v)v\partial_xf(x,v,t)\,dv\,dx= \\
		&\frac{1}{2}\int_{\R}\int_{\R}\int_0^1\int_0^1B(x_\ast-x)(\phi(x,v')-\phi(x,v))f(x,v,t)f(x_\ast,v_\ast,t)\,dv\,dv_\ast\,dx\,dx_\ast \\
		&+\epsilon\int_{\R}\int_0^1(\phi(x,v'')-\phi(x,v))f(x,v,t)\,dv\,dx,
	\end{aligned}
	\label{eq:kinetic}
\end{align}
where $f=f(x,v,t):\R\times [0,\,1]\times (0,\,+\infty)\to\R_+$ is the distribution function of the microscopic state $(x,v)$ of the vehicles at time $t$. In essence, $f(x,v,t)\,dx\,dv$ gives the probability that a vehicle has a position comprised between $x$ and $x+dx$ and a speed comprised between $v$ and $v+dv$ at time $t$. On the right-hand side of~\eqref{eq:kinetic} we have denoted by
\begin{subequations}
\begin{align}
	& v'=v+\lambda(h(\rho))(v_\ast-v) \label{eq:int.FTL} \\
	& v''=v+a(V(h(\rho))-v) \label{eq:int.OV}
\end{align}
\end{subequations}
the speed update rules based on FTL interactions and OV relaxation, respectively.

Two remarks are in order about the first term on the right-hand side of~\eqref{eq:kinetic}. First, this term is written under the (usual) assumption of \textit{propagation of chaos}, which states that any two randomly chosen vehicles are statistically independent at the moment of their interaction. This allows one to write the joint distribution function $f_2=f_2(x,x_\ast,v,v_\ast,t)$ of the interacting vehicles as the product of the two marginal distributions:
$$ f_2(x,x_\ast,v,v_\ast,t)=f(x,v,t)f(x_\ast,v_\ast,t), $$
where $(x,v)$ and $(x_\ast,v_\ast)$ are the position-speed pairs of the interacting vehicles. Second, this term has the form of a \textit{Povzner collisional operator}~\cite{fornasier2011PHYSD,povzner1962AMSTS}, $B$ being a collision kernel which averages the positions of the interacting vehicles.

We observe that FTL interactions are pointwise in space, namely vehicles in $x$ interact with vehicles in a precise other position $x_\ast$. This suggests two possible choices for the collision kernel $B$. The perhaps most standard one is $B(x_\ast-x)=\delta(x_\ast-x)$, where $\delta$ denotes the Dirac delta, meaning that vehicles interact if $x_\ast=x$, hence if they occupy the same position. This makes~\eqref{eq:kinetic} a \textit{Boltzmann-type} kinetic equation, which however has proved to be ineffective for appropriately modelling traffic dynamics. Indeed, the assumption of localised interactions, combined with the non-negativity of the vehicle speed, prevents this equation from describing the backward propagation of traffic waves, cf.~\cite{klar1997JSP}. An alternative choice is instead $B(x_\ast-x)=\delta(x_\ast-x-h(\rho))$, implying that $x_\ast=x+h(\rho)$ and thus that the second vehicle of the interacting pair leads the first one. This makes~\eqref{eq:kinetic} an \textit{Enskog-type} kinetic equation, which yields instead more consistent aggregate traffic descriptions, cf.~\cite{dimarco2020JSP,klar1997JSP}.

Setting therefore
$$ B(x_\ast-x)=\delta(x_\ast-x-h(\rho)) $$
and choosing the observable $\phi$ of the form $\phi(x,v)=\psi(x)\varphi(v)$ we rewrite~\eqref{eq:kinetic} in strong form with respect to $x$ and $t$ by invoking the arbitrariness of $\psi$:
\begin{align}
	\begin{aligned}[b]
		\partial_t\int_0^1\varphi(v)f(x,v,t)\,dv\,dx+&\partial_x\int_0^1\varphi(v)vf(x,v,t)\,dv= \\
		&\frac{1}{2}\int_0^1\int_0^1(\varphi(v')-\varphi(v))f(x,v,t)f(x+h(\rho),v_\ast,t)\,dv\,dv_\ast \\
		&+\epsilon\int_0^1(\varphi(v'')-\varphi(v))f(x,v,t)\,dv.
	\end{aligned}
	\label{eq:Enskog}
\end{align}
Moreover, with the first order approximation
\begin{equation}
	f(x+h(\rho),v_\ast,t)\approx f(x,v_\ast,t)+\partial_xf(x,v_\ast,t)h(\rho)
	\label{eq:approx.f}
\end{equation}
we further specialise it as
\begin{align}
	\begin{aligned}[b]
		\partial_t\int_0^1\varphi(v)f(x,v,t)\,dv\,dx+&\partial_x\int_0^1\varphi(v)vf(x,v,t)\,dv= \\
		&\frac{1}{2}\int_0^1\int_0^1(\varphi(v')-\varphi(v))f(x,v,t)f(x,v_\ast,t)\,dv\,dv_\ast \\
		&+\frac{h(\rho)}{2}\int_0^1\int_0^1(\varphi(v')-\varphi(v))f(x,v,t)\partial_xf(x,v_\ast,t)\,dv\,dv_\ast \\
		&+\epsilon\int_0^1(\varphi(v'')-\varphi(v))f(x,v,t)\,dv.
	\end{aligned}
	\label{eq:Enskog.approx}
\end{align}
By introducing the bilinear collisional operator $Q(f,g)$, whose action against an observable quantity $\varphi(v)$ is defined as
$$ (Q(f,g),\,\varphi):=\frac{1}{2}\int_0^1\int_0^1(\varphi(v')-\varphi(v))f(x,v,t)g(x,v_\ast,t)\,dv\,dv_\ast, $$
and the linear relaxation operator $R(f)$ such that
$$ (R(f),\,\varphi):=\int_0^1(\varphi(v'')-\varphi(v))f(x,v,t)\,dv, $$
we formally restate~\eqref{eq:Enskog.approx} in the strong form
\begin{equation}
	\partial_tf+v\partial_xf=Q(f,f)+h(\rho)Q(f,\partial_xf)+\epsilon R(f).
	\label{eq:Enskog.strong}
\end{equation}
In particular, we recognise that $Q$ is a classical Boltzmann-type collisional operator and that, under the approximation~\eqref{eq:approx.f}, the original Enskog-type collisional term is rewritten as the superposition of local interactions expressed by $Q(f,f)$ and a correction given by $Q(f,\partial_xf)$.

\begin{remark}[About the interaction rules~\eqref{eq:int.FTL}-\eqref{eq:int.OV}]
The traffic density $\rho$ appearing in the interaction rules~\eqref{eq:int.FTL}-\eqref{eq:int.OV} is naturally expressed in terms of the distribution function $f$ as
$$ \rho(x,t):=\int_0^1f(x,v,t)\,dv. $$

Moreover, we observe that for physical consistency the interaction rules~\eqref{eq:int.FTL}-\eqref{eq:int.OV} have to guarantee $v',\,v''\in [0,\,1]$ for all $v,\,v_\ast\in [0,\,1]$. In~\eqref{eq:int.FTL} this is readily obtained if the function $\lambda$ is such that $0\leq\lambda(h)\leq 1$ for all $h\in\R_+$, for then $v'$ is a convex combination of $v$ and $v_\ast$. Notice that function~\eqref{eq:lambda.bounded} fulfills this requirement for $0\leq\lambda_0\leq 1$. In~\eqref{eq:int.OV} the analogous condition is $0\leq a\leq 1$, which makes $v''$ a convex combination of $v$ and $V(h)$.
\label{rem:phys.cons}
\end{remark}

\subsection{Hydrodynamic limit}
Since the parameter $\epsilon$ is small by assumption, we may use it as a small \textit{Knudsen number} to pass from the kinetic description~\eqref{eq:Enskog.strong} to the hydrodynamic regime. To this purpose, we introduce the following hyperbolic scaling of time and space:
\begin{equation}
	t\to\frac{t}{\epsilon}, \qquad x\to\frac{x}{\epsilon},
	\label{eq:hyp_scal}
\end{equation}
whence $\partial_tf\to\epsilon\partial_tf$ and $\partial_xf\to\epsilon\partial_xf$. Since $Q$ is bilinear we also have $Q(f,\epsilon\partial_xf)=\epsilon Q(f,\partial_xf)$, so that finally we rewrite~\eqref{eq:Enskog.strong} as
$$ \partial_tf+v\partial_xf=\frac{1}{\epsilon}Q(f,f)+h(\rho)Q(f,\partial_xf)+R(f). $$
In the hydrodynamic limit $\epsilon\to 0^+$ we may split this equation in two contributions:
\begin{subequations}
\begin{enumerate}[label=(\roman*)]
\item the \textit{local} equilibrium of FTL interactions expressed by
\begin{equation}
	Q(f,f)=0,
	\label{eq:split-local.int}
\end{equation}
which returns an equilibrium speed distribution at time $t$ and in the point $x$ on the hydrodynamic scale;
\item the transport of such a local equilibrium distribution on the hydrodynamic scale:
\begin{equation}
	\partial_tf+v\partial_xf=h(\rho)Q(f,\partial_xf)+R(f),
	\label{eq:split-transport}
\end{equation}
which also includes the first-order-in-space correction of the FTL interactions and the OV relaxation dynamics.
\end{enumerate}
\end{subequations}

We now study in more detail each of these contributions.

\subsubsection{Local Maxwellian}
\label{sect:local_Maxwellian}
To study the local equilibrium it is convenient to rewrite~\eqref{eq:split-local.int} in evolutionary form on the quick time scale, say $\tau$, of the local microscopic interactions. Therefore we consider the equation
$$ \partial_\tau f=Q(f,f) $$
and we study its asymptotic trend for $\tau\to +\infty$, which corresponds to the equilibrium instantaneously reached at time $t$ on the hydrodynamic scale. Using the definition of $Q$, in weak form we have
$$ \partial_\tau\int_0^1\varphi(v)f(x,v,\tau)\,dv=\frac{1}{2}\int_0^1\int_0^1(\varphi(v')-\varphi(v))f(x,v,\tau)f(x,v_\ast,\tau)\,dv\,dv_\ast, $$
where $x\in\R$ acts here as a parameter and $v'$ is given by~\eqref{eq:int.FTL}.

By choosing $\varphi(v)=1$ we readily obtain $\partial_\tau\rho=0$, meaning that the traffic density is conserved by the local FTL interactions. This traffic density coincides therefore with the one in the point $x$ at time $t$ on the hydrodynamic scale, i.e. $\rho(x,t)$.

Next, by choosing $\varphi(v)=v$ and defining the \textit{mean speed}
$$ u(x,\tau):=\frac{1}{\rho(x,\tau)}\int_0^1vf(x,v,\tau)\,dv $$
we see that $\partial_\tau(\rho u)=0$ because
\begin{align*}
	(Q(f,f),v) &= \frac{1}{2}\int_0^1\int_0^1(v'-v)f(x,v,\tau)f(x,v_\ast,\tau)\,dv\,dv_\ast \\
	&= \frac{\lambda(h(\rho))}{2}\int_0^1\int_0^1(v_\ast-v)f(x,v,\tau)f(x,v_\ast,\tau)\,dv\,dv_\ast \\
	&= 0,
\end{align*}
which implies that also the mean speed $u$ is conserved by the local FTL interactions. It coincides then with the one in the point $x$ at time $t$ on the hydrodynamic scale, i.e. $u(x,t)$.

These conservations imply that the local equilibrium speed distribution resulting from~\eqref{eq:split-local.int} is parametrised by the quantities $\rho$, $u$. We denote such a distribution as $M_{\rho,u}=M_{\rho,u}(v)$, the letter $M$ recalling that it is the analogous of the \textit{local Maxwellian} of the classical kinetic theory. In particular, we have the following relationships:
$$ \int_0^1M_{\rho,u}(v)\,dv=\rho(x,t), \qquad \frac{1}{\rho(x,t)}\int_0^1vM_{\rho,u}(v)\,dv=u(x,t). $$

To gain further insights into the local Maxwellian $M_{\rho,u}$ it is useful to investigate the \textit{total energy} of the distribution $f$:
$$ E(x,\tau):=\frac{1}{\rho(x,\tau)}\int_0^1v^2f(x,v,\tau)\,dv. $$
By computing its evolution with $\varphi(v)=v^2$ we discover, for $\rho>0$:
\begin{align*}
	\partial_\tau E &= \frac{1}{\rho}(Q(f,f),v^2) \\
	&= \frac{1}{2\rho}\int_0^1\int_0^1\left((v')^2-v^2\right)f(x,v,\tau)f(x,v_\ast,\tau)\,dv\,dv_\ast \\
	&= \lambda(h(\rho))\bigl(1-\lambda(h(\rho))\bigr)\rho(u^2-E).
\end{align*}
Since $\lambda$ is typically a decreasing function of the headway $h$, i.e. an increasing function of the density $\rho$ (cf. the examples given at the beginning of Section~\ref{sect:kinetic}), from Remark~\ref{rem:phys.cons} we infer $0<\lambda(h(\rho))<1$ for $0<\rho<+\infty$. Then $E\to u^2$ for $\tau\to +\infty$, which implies that the total energy of the local Maxwellian is $u^2$. Consequently its variance is zero and $M_{\rho,u}$ is fully characterised as
\begin{equation}
	M_{\rho,u}(v)=\rho\delta(v-u).
	\label{eq:M}
\end{equation}
The local FTL interactions cause therefore the emergence of a local \textit{group velocity} $u$ of the vehicles, which will then evolve on the hydrodynamic scale.

\subsubsection{Macroscopic equations}
Macroscopic equations are obtained by plugging the local Maxwellian~\eqref{eq:M} into~\eqref{eq:split-transport} to determine the evolution of the hydrodynamic parameters $\rho$, $u$. In weak form,~\eqref{eq:split-transport} with $f=M_{\rho,u}$ reads
\begin{multline*}
	\partial_t\int_0^1\varphi(v)M_{\rho,u}(v)\,dv+\partial_x\int_0^1\varphi(v)vM_{\rho,u}(v)\,dv \\
	=\frac{h(\rho)}{2}\int_0^1\int_0^1(\varphi(v')-\varphi(v))M_{\rho,u}(v)\partial_xM_{\rho,u}(v_\ast)\,dv\,dv_\ast
		+\int_0^1(\varphi(v'')-\varphi(v))M_{\rho,u}(v)\,dv,
\end{multline*}
where $v'$, $v''$ are given by~\eqref{eq:int.FTL},~\eqref{eq:int.OV}, respectively. With~\eqref{eq:M} this equation specialises as
$$ \partial_t(\rho\varphi(u))+\partial_x(\rho u\varphi(u))=\frac{\rho^2\lambda(h(\rho))h(\rho)}{2}\varphi'(u)\partial_xu+\rho\left[\varphi(u+a(V(h(\rho))-u))-\varphi(u)\right] $$
whence, using the collisional invariants $\varphi(v)=1,\,v$, we get the system of macroscopic equations
\begin{equation*}
	\begin{cases}
		\partial_t\rho+\partial_x(\rho u)=0 \\
		\partial_t(\rho u)+\partial_x(\rho u^2)=\dfrac{\rho^2\lambda(h(\rho))h(\rho)}{2}\partial_xu+\rho a(V(h(\rho))-u).
	\end{cases}
\end{equation*}
Defining a (pseudo) \textit{traffic pressure} $p=p(\rho)$ via the relationship
\begin{equation}
	p'(\rho):=\frac{\lambda(h(\rho))h(\rho)}{2},
	\label{eq:p}
\end{equation}
where $'$ stands here for the derivative with respect to $\rho$, and using the continuity equation (first equation of the system) we may rewrite on the whole the macroscopic model as
\begin{equation}
	\begin{cases}
		\partial_t\rho+\partial_x(\rho u)=0 \\
		\partial_t(u+p(\rho))+u\partial_x(u+p(\rho))=a(V(h(\rho))-u),
	\end{cases}
	\label{eq:inhom_ARZ}
\end{equation}
whence we recognise an \textit{inhomogeneous ARZ} traffic model~\cite{greenberg2001SIAP,rascle2002MCM}. From~\eqref{eq:p} we see precisely how the traffic pressure depends on the sensitivity and the headway of the drivers. In particular, owing to Assumption~\ref{ass:Vh} the traffic pressure is a non-decreasing function of the traffic density, which correctly reproduces the anticipatory behaviour of the drivers.

\subsection{A Bando-type stability condition of the uniform flow}
\label{sect:stability}
In~\cite{bando1995PRE} Bando and co-authors showed that the microscopic OV model~\eqref{eq:OV} predicts the stability of the uniform traffic flow, i.e. the one in which all vehicles are at a constant distance $h_0\geq 0$ from each other, if the following condition is met:
\begin{equation}
	V'(h_0)<\frac{a}{2},
	\label{eq:Bando}
\end{equation}
which relates the variation of the optimal speed near the uniform headway $h_0$ to the relaxation rate $a$. Here we show that model~\eqref{eq:p}-\eqref{eq:inhom_ARZ} exhibits a similar stability condition at the macroscopic scale, in which however the important role is played by the sensitivity $\lambda$ rather than by $a$.

At the macroscopic scale, a uniform traffic flow corresponds to constant density and mean speed~\cite{flynn2009PRE,kerner1993PRE}. Let us consider then a constant solution to~\eqref{eq:inhom_ARZ}, namely
$$ \rho=\rho_0>0, \qquad u=V(h_0), $$
where $h_0:=h(\rho_0)$, and a related perturbation
\begin{equation}
	\rho(x,t)=\rho_0+\eta\hat{\rho}(x,t), \qquad u(x,t)=V(h_0)+\eta\hat{u}(x,t),
	\label{eq:perturbations}
\end{equation}
where $\eta>0$ is a small parameter. Plugging into~\eqref{eq:inhom_ARZ} and retaining only the terms up to the first order in $\eta$ we get the following linearised system of equations for the perturbations $\hat{\rho}$, $\hat{u}$:
\begin{equation}
	\begin{cases}
		\partial_t\hat{\rho}+\rho_0\partial_x\hat{u}+V(h_0)\partial_x\hat{\rho}=0 \\[2mm]
		\partial_t\hat{u}+\left(V(h_0)-\dfrac{\rho_0\lambda(h_0)h_0}{2}\right)\partial_x\hat{u}=a\bigl(V'(h_0)h'(\rho_0)\hat{\rho}-\hat{u}\bigr).
	\end{cases}
	\label{eq:lin_sys}
\end{equation}

To study the linear stability of the null configuration $\hat{\rho}=\hat{u}=(0,0)$, which corresponds to the linear stability of the constant solution $(\rho,u)=(\rho_0,V(h_0))$, it is convenient to consider~\eqref{eq:lin_sys} in a bounded domain with periodic boundary conditions. Letting $L>0$ be the length of the spatial domain, we may then look for solutions of the form
$$ \hat{\rho}(x,t)=\hat{\rho}_0e^{i\frac{2\pi}{L}kx+\omega t}, \qquad \hat{u}(x,t)=\hat{u}_0e^{i\frac{2\pi}{L}kx+\omega t}, $$
where $i$ is the imaginary unit, $k\in\mathbb{Z}$ is the wavenumber and $\omega\in\mathbb{C}$ is the frequency of the perturbation. Moreover, $\hat{\rho}_0,\,\hat{u}_0\in\mathbb{C}$ are the initial amplitudes of the perturbations. Plugging into~\eqref{eq:lin_sys} yields
\begin{equation*}
	\begin{pmatrix}
		\omega+iV(h_0)\frac{2\pi}{L}k & i\rho_0\frac{2\pi}{L}k \\[2mm]
		-aV'(h_0)h'(\rho_0) & \omega+i\left(V(h_0)-\frac{\rho_0\lambda(h_0)h_0}{2}\right)\frac{2\pi}{L}k+a
	\end{pmatrix}
	\begin{pmatrix}
		\hat{\rho}_0 \\
		\hat{u}_0
	\end{pmatrix}
	=
	\begin{pmatrix}
		0 \\
		0
	\end{pmatrix},
\end{equation*}
which admits non-trivial solutions $(\hat{\rho}_0,\hat{u}_0)\neq (0,0)$ provided the determinant of the coefficient matrix vanishes. This produces the following dispersion relations:
$$ \omega^\pm(k)=-\frac{a}{2}(1\pm\Omega(k))-i\left(V(h_0)-\frac{\rho_0\lambda(h_0)h_0}{4}\right)\frac{2\pi}{L}k, $$
where we have set
$$ \Omega(k):=\sqrt{1-\frac{\rho_0^2\lambda^2(h_0)h_0^2}{a^2L^2}\pi^2k^2-2i\rho_0\frac{4V'(h_0)h'(\rho_0)+\lambda(h_0)h_0}{aL}\pi k}. $$
Stability is obtained if $\Real{\omega^\pm}\leq 0$, i.e. if $1\pm\Real{\Omega(k)}\geq 0$ and finally $\vert\Real{\Omega(k)}\vert\leq 1$. After some algebraic manipulations, this leads to the condition
$$ \frac{32\pi^2\rho_0^2V'(h_0)h'(\rho_0)}{a^2L^2}\Bigl(2V'(h_0)h'(\rho_0)+\lambda(h_0)h_0\Bigr)k^2\leq 0, $$
which, considering that $V'(h_0)>0$ and $h'(\rho_0)<0$ because of Assumption~\ref{ass:Vh}, gives the (linear) stability to all waves if
\begin{equation}
	V'(h_0)\leq\frac{\lambda(h_0)}{2}\cdot\frac{h_0}{\vert h'(\rho_0)\vert}.
	\label{eq:stability}
\end{equation}
If~\eqref{eq:stability} is fulfilled with strict inequality then the uniform traffic state is also (linearly) attractive, i.e. small perturbations dissipate exponentially fast in time.

\begin{remark}
Recalling~\eqref{eq:p}, condition~\eqref{eq:stability} may be rewritten as
$$ \left\vert\left.\frac{d}{d\rho}V(h(\rho))\right\vert_{\rho_0}\right\vert\leq p'(\rho_0), $$
which establishes a direct relationship between the derivatives of the optimal speed and the traffic pressure computed at the equilibrium density $\rho_0$.
\end{remark}

Compared to~\eqref{eq:Bando}, we notice that this condition does not involve the relaxation rate $a$. Therefore we may assert that the superposition of FTL and OV microscopic dynamics, \textit{under the key assumption that the latter are less frequent than the former}, produces a macroscopic flow whose stability is essentially ruled by the FTL driver sensitivity.

It is straightforward to check that our condition~\eqref{eq:stability} is consistent with the stability condition derived in~\cite{ramadan2021SEMAI-SIMAI} for a generic inhomogeneous ARZ model, in which however the traffic pressure is not precisely related to the sensitivity and the headway of the drivers.

\section{Fast OV dynamics}
\label{sect:fast_OV}
We now reconsider the assumption of different frequency between FTL and OV speed updates made in Section~\ref{sect:kinetic}. In particular, we analyse the case in which the two processes take place at the same rate, i.e. both $\Theta$ and $\Xi$ in~\eqref{eq:particle_model} are Bernoulli random variables with the same parameter $\Delta{t}$:
\begin{equation}
	\Theta,\,\Xi\sim\operatorname{Bernoulli}(\Delta{t}).
	\label{eq:Bernoulli.fast_OV}
\end{equation}

The kinetic equation~\eqref{eq:Enskog.strong} takes then the form
$$ \partial_tf+v\partial_xf=Q(f,f)+h(\rho)Q(f,\partial_xf)+R(f), $$
which, after the hyperbolic scaling~\eqref{eq:hyp_scal} of space and time, leads to
$$ \partial_tf+v\partial_xf=\frac{1}{\epsilon}(Q(f,f)+R(f))+h(\rho)Q(f,\partial_xf) $$
and hence, in the hydrodynamic limit $\epsilon\to 0^+$, to the splitting
\begin{subequations}
\begin{align}
	& Q(f,f)+R(f)=0 \label{eq:split.fast-local.int} \\
	& \partial_tf+v\partial_xf=h(\rho)Q(f,\partial_xf). \label{eq:split.fast-transport}
\end{align}
\end{subequations}

From~\eqref{eq:split.fast-local.int} we see that the local Maxwellian is now produced by the superposition of local FTL interactions expressed by the operator $Q$ and the OV relaxation process expressed by the operator $R$. To study the trend towards the local equilibrium we proceed like in Section~\ref{sect:local_Maxwellian}, i.e. we consider the evolutionary equation
$$ \partial_\tau f=Q(f,f)+R(f) $$
on the time scale $\tau$ of the quick local interactions and we investigate its behaviour for $\tau\to +\infty$. The weak form of this equation is
\begin{align*}
	\partial_\tau\int_0^1\varphi(v)f(x,v,\tau)\,dv &= \frac{1}{2}\int_0^1\int_0^1(\varphi(v')-\varphi(v))f(x,v,\tau)f(x,v_\ast,\tau)\,dv\,dv_\ast \\
	&\phantom{=} +\int_0^1(\varphi(v'')-\varphi(v))f(x,v,\tau)\,dv,
\end{align*}
where $v'$, $v''$ are like in~\eqref{eq:int.FTL},~\eqref{eq:int.OV}, respectively. For $\varphi(v)=1$ we deduce again $\partial_\tau\rho=0$, i.e. the local conservation in time of the traffic density. For $\varphi(v)=v$ we obtain instead, with $\rho>0$,
$$ \partial_\tau u=a(V(h(\rho))-u), $$
which for $a>0$ says that the local mean speed converges exponentially fast at rate $a$ to the optimal speed $V(h(\rho))$. Since the only quantity conserved by the local interactions is the traffic density, we expect a local Maxwellian $M_\rho=M_\rho(v)$ parametrised only by $\rho$.

To investigate in more detail the structure of $M_\rho$ we consider the trend of the total energy $E$, which we obtain for $\varphi(v)=v^2$. In particular, for $\rho>0$ it results
$$ \partial_\tau E=\mu_\rho(g_\rho(u)-E), $$
where
$$ \mu_\rho:=\rho\lambda(h(\rho))[1-\lambda(h(\rho))]+a(2-a) $$
and
$$ g_\rho(u):=\frac{\rho\lambda(h(\rho))[1-\lambda(h(\rho))]u^2+2aV(h(\rho))u+a^2[V(h(\rho))-2u]V(h(\rho))}{\mu_\rho}. $$
Owing to Remark~\ref{rem:phys.cons}, we observe that $\mu_\rho>0$ whenever $a>0$. Moreover, since $u\to V(h(\rho))$ exponentially fast for $\tau\to +\infty$ we also deduce
$$ g_\rho(u)\to V^2(h(\rho)) $$
exponentially fast, thus we conclude $E\to V^2(h(\rho))$ for $\tau\to +\infty$. The local Maxwellian has then null variance, hence it is fully characterised as
$$ M_\rho(v)=\rho\delta(v-V(h(\rho))). $$

Plugging it in~\eqref{eq:split.fast-transport} we obtain a single conservation law for the evolution of the traffic density:
\begin{equation}
	\partial_t\rho+\partial_x(\rho V(h(\rho)))=0,
	\label{eq:LWR}
\end{equation}
namely a Lighthill-Whitham-Richards (LWR) traffic model, which may be formally viewed as the limit of~\eqref{eq:inhom_ARZ} in the relaxation-dominated regime $a\to +\infty$. Hence, \textit{a high relaxation rate in~\eqref{eq:inhom_ARZ} is the macroscopic counterpart of an OV frequency comparable with (notice: not necessarily much higher than) the FTL frequency at the particle level}.

Clearly, in~\eqref{eq:LWR} every uniform traffic state $\rho=\rho_0$ is (linearly) stable but not attractive, i.e. small perturbations do not dissipate. Nevertheless, the LWR model~\eqref{eq:LWR} plays a role in determining the stability properties of the inhomogeneous ARZ model~\eqref{eq:inhom_ARZ}. Indeed, according to the so-called \textit{sub-characteristic condition}~\cite{seibold2013NHM,whitham1959CPAM} the uniform traffic state $U_0:=(\rho_0,V(h_0))$ is (linearly) stable for model~\eqref{eq:inhom_ARZ} if the characteristic speed of~\eqref{eq:LWR} computed at $U_0$ lies between the two characteristic speeds of~\eqref{eq:inhom_ARZ} computed in turn at $U_0$. In formulas:
$$ V(h_0)-\rho_0p'(\rho_0)\leq V(h_0)+\rho_0V'(h_0)h'(\rho_0)\leq V(h_0), $$
which can be readily checked to lead to~\eqref{eq:stability}.

\section{Numerical tests}
\label{sect:numerics}
In this section, we perform some numerical tests aimed at supporting the theoretical findings of the previous sections. First, we compare the numerical solutions produced by the particle model~\eqref{eq:particle_model} in the two regimes considered in Sections~\ref{sect:slow_OV},~\ref{sect:fast_OV} with the solutions of the ARZ and LWR macroscopic models~\eqref{eq:p}-\eqref{eq:inhom_ARZ},~\eqref{eq:LWR}. This allows us to test also numerically the convergence of the particle model to either macroscopic model depending on the scaling parameter $\epsilon$. Second, we test numerically the stability condition~\eqref{eq:stability} for the inhomogeneous ARZ model~\eqref{eq:inhom_ARZ}.

\subsection{Convergence of the particle system to the macroscopic models}
Let $N\in\mathbb{N}$ be the total number of particles used in the particle model and let $\{(x_i^n,\,v_i^n)\}_{i=1,\dots,N}$ be their position-speed pairs at time $t^n:=n\Delta{t}$, $n=0,\,1,\,2,\,\dots$, where $\Delta{t}>0$ is a fixed time step. We take as spatial domain the interval $[-1,\,1]$ with periodic boundary conditions, thus $x_i^n\in [-1,\,1]$ for all $i=1,\,\dots,\,N$ and all $n$. Moreover, $v_i^n\in [0,\,1]$ consistently with the non-dimensional speed variable introduced in Section~\ref{sect:slow_OV}. We denote by
$$ f_N^n(x,v):=\frac{1}{N}\sum_{i=1}^{N}\delta(x-x_i^n)\otimes\delta(v-v_i^n) $$
the empirical particle distribution at time $t^n$.

\subsubsection{Computing macroscopic quantities out of the particles}
To compute the macroscopic quantities of interest we partition the space interval in cells $E_j:=[x_j-\frac{\Delta{x}}{2},\,x_j+\frac{\Delta{x}}{2}]$ of constant size $\Delta{x}>0$ centred in the points $x_j$, $j=1,\,\dots,\,J$, where $J\in\mathbb{N}$ is the total number of cells. Thus $\Delta{x}=\frac{2}{J}$. Next we introduce the following piecewise constant approximation of the macroscopic particle density:
$$ \rho^n(x):=\sum_{j=1}^{J}\rho^n_j\chi_{E_j}(x), $$
where $\chi_{E_j}$ is the characteristic function of cell $E_j$, and we impose
$$ \int_{E_j}\rho^n(x)\,dx=\int_{E_j}\int_0^1f^n_N(x,v)\,dv\,dx, \qquad \forall\,j=1,\,\dots,\,J, $$
which yields
\begin{equation}
	\rho^n_j=\frac{N_j^n}{N\Delta{x}} \qquad \text{with} \qquad N_j^n:=\#\{i=1,\,\dots,\,N\,:\,x_i^n\in E_j\}.
	\label{eq:rho^n_j}
\end{equation}
Similarly, we introduce the following piecewise constant approximation of the mean speed of the particles:
$$ u^n(x):=\sum_{j=1}^{J}u^n_j\chi_{E_j}(x) $$
and we impose
$$ \int_{E_j}\rho^n(x)u^n(x)\,dx=\int_{E_j}\int_0^1vf_N^n(x,v)\,dv\,dx, \qquad \forall\,j=1,\,\dots,\,J, $$
which, together with~\eqref{eq:rho^n_j}, gives
$$ u^n_j=\frac{1}{N^n_j}\sum_{i\,:\,x_i^n\in E_j}v_i^n, $$
i.e. the arithmetic mean of the speeds of the particles in cell $E_j$ at time $n$.

Summarising, at each time step we build the macroscopic quantities of interest out of the particles as
\begin{equation}
	\rho^n(x)=\frac{1}{N\Delta{x}}\sum_{j=1}^{J}N^n_j\chi_{E_j}(x), \qquad
		u^n(x)=\sum_{j=1}^{J}\frac{1}{N^n_j}\left(\sum_{i\,:\,x_i^n\in E_j}v^n_i\right)\chi_{E_j}(x).
	\label{eq:rho^n_u^n}
\end{equation}
	
\subsubsection{Prescribing initial conditions}
\label{sect:initcond}
As far as initial conditions are concerned, it is more practical to prescribe them at the macroscopic scale and then translate them at the particle level. In particular, we consider
\begin{equation}
	\rho(x,0)=
	\begin{cases}
		\rho_\ell & \text{for } x<0 \\
		\rho_r & \text{for } x>0,
	\end{cases}
	\qquad
	u(x,0)=
	\begin{cases}
		u_\ell & \text{for } x<0 \\
		u_r & \text{for } x>0
	\end{cases}
	\label{eq:initcond}
\end{equation}
with $\rho_\ell,\,\rho_r,\,u_\ell,\,u_r\in [0,\,1]$, which define a classical Riemann problem. The point is now how to generate the pairs $\{(x_i^0,\,v_i^0)\}_{i=1,\dots,N}$ so as to match~\eqref{eq:initcond} at the particle level.

Using the first formula in~\eqref{eq:rho^n_u^n} and integrating for $x\in [-1,\,0]$ we discover
$$ \int_{-1}^{0}\rho(x,0)\,dx=\int_{-1}^{0}\rho^0(x)\,dx \qquad \text{i.e.} \qquad
	\rho_\ell=\frac{1}{N}\sum_{j\,:\,E_j\subseteq [-1,\,0]}N_j^0. $$
Next, considering that $N^0_\ell:=\sum_{j\,:\,E_j\subseteq [-1,\,0]}N_j^0$ is the total number of particles initially distributed in $[-1,\,0]$, we deduce $N^0_\ell=N\rho_\ell$. Therefore, to mimic the constant density $\rho_\ell$ for $x<0$ it is sufficient to sample $N\rho_\ell$ particle positions $x^0_i$ from a uniform distribution supported in $[-1,\,0]$. Analogously, to reproduce the constant density $\rho_r$ for $x>0$ it is sufficient to sample $N^0_r:=N\rho_r$ particle positions $x^0_i$ from a uniform distribution supported in $[0,\,1]$. Notice that $N^0_\ell+N^0_r=N(\rho_\ell+\rho_r)$, therefore the condition
$$ \rho_\ell+\rho_r=1 $$
is necessary for consistency. From a different point of view, this corresponds to the requirement that the particle distribution function be a probability density.

Finally, to generate the speeds of the particles consistently with~\eqref{eq:initcond} we sample $N^0_\ell$ values $v^0_i$ from a uniform distribution supported in $[0,\,2u_\ell]$, hence with mean $u_\ell$, for the particles located in $[-1,\,0]$ and other $N^0_r$ values $v^0_i$ from a uniform distribution supported in $[0,\,2u_r]$, hence with mean $u_r$, for the particles located in $[0,\,1]$.

\begin{figure}[!t]
\centering
\includegraphics[width=0.33\textwidth]{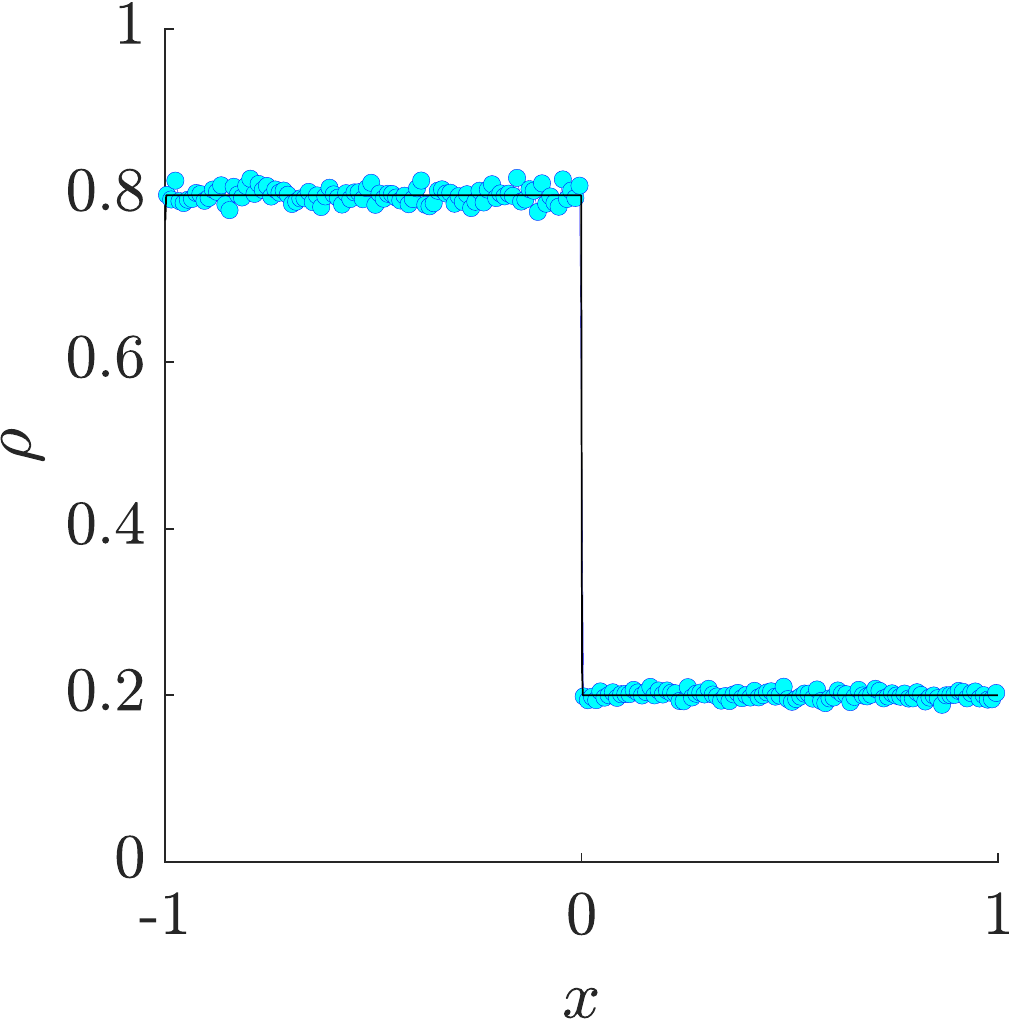}
\includegraphics[width=0.33\textwidth]{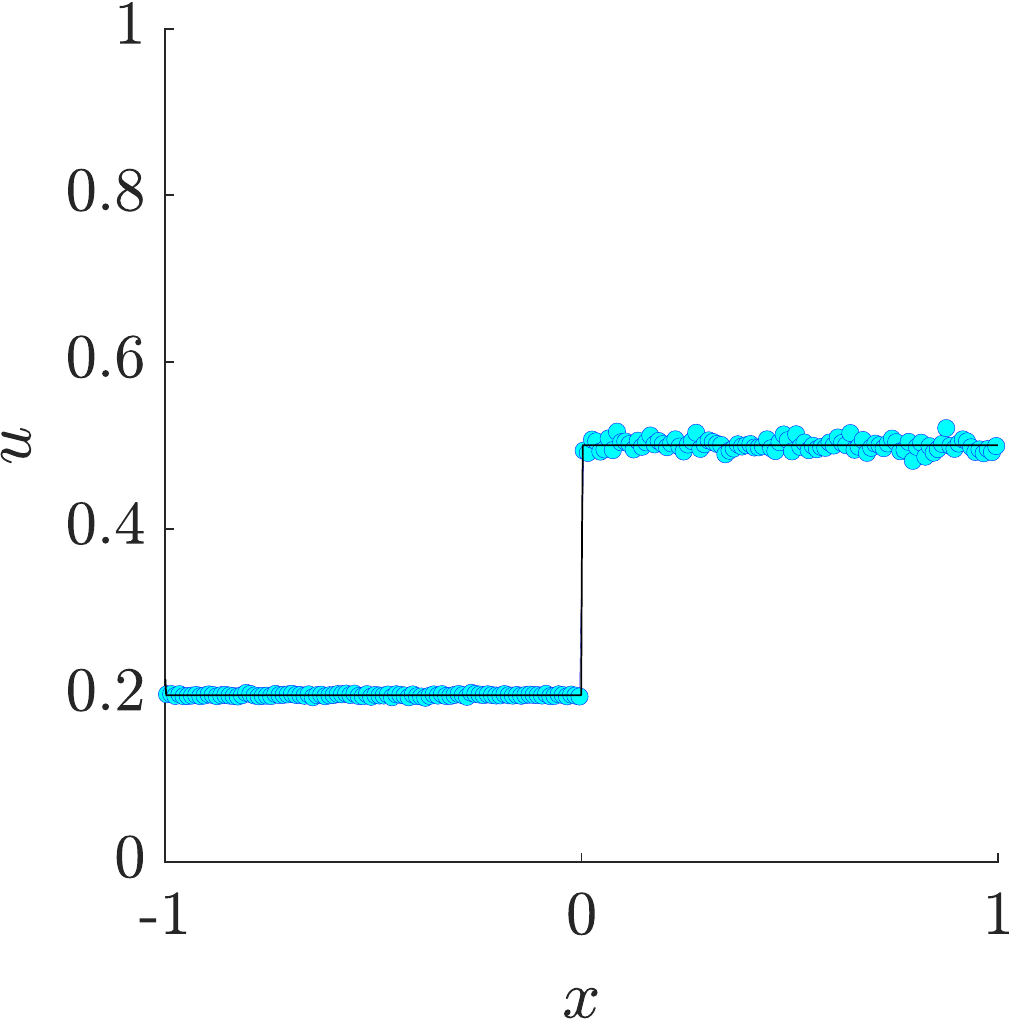}
\caption{Initial density (left) and mean speed (right) generated at the particle level (markers) from the macroscopic values $\rho_\ell=0.8$, $\rho_r=0.2$, $u_\ell=0.2$, $u_r=0.5$ (solid line, cf.~\eqref{eq:initcond}) as described in Section~\ref{sect:initcond}.}
\label{fig:initcond}
\end{figure}

Figure~\ref{fig:initcond} shows the initial density and mean speed computed out of particle positions and speeds generated with this procedure, starting from a prescribed macroscopic initial condition of the form~\eqref{eq:initcond}.

\subsubsection{Particle algorithm and comparison with macroscopic results}
\begin{algorithm}[!t]
	\caption{Monte Carlo algorithm for the particle model~\eqref{eq:particle_model}}
	\label{alg:MC}
	Fix $\Delta{t}\leq\epsilon$\; \label{alg:Dt}
	Fix $\Delta{x}=O(h)$\; \label{alg:Dx}
	\For{$n=0,\,1,\,2,\,\dots$}{
		Compute $\rho^n(x)$, $u^n(x)$ as in~\eqref{eq:rho^n_u^n}\;
		\For{$j=1,\,\dots,\,J$}{
			Find the particles belonging to the cell $E_j$. Let $\mathcal{E}_j:=\{(x^n_i,\,v^n_i)\,:\,x^n_i\in E_j\}$\;
			\Repeat{no particles are left in $\mathcal{E}_j$ (if $\#\mathcal{E}_j$ is odd then discard randomly one particle)}{
				Pick randomly and without replacement $(x^n_h,\,v^n_h),\,(x^n_k,\,v^n_k)\in\mathcal{E}_j$ ($h\neq k$)\; \label{alg:pick_particles}
				Let $x^n_h\leq x^n_k$, otherwise switch the two particles\; \label{alg:switch}
				Sample $\Theta=\theta\in\{0,\,1\}$\; \label{alg:Bernoulli.Theta}
				Update $v^n_h$ to $v^{n+1/2}_h$ due to an FTL interaction: \newline{} \label{alg:FTL}
					$v^{n+1/2}_h=v^n_h+\theta\lambda(h(\rho^n_j))(v^n_k-v^n_h)$\;
				Sample two independent values $\Xi=\xi_h,\,\xi_k\in\{0,\,1\}$\; \label{alg:Bernoulli.Xi}
				Update $v^{n+1/2}_h,\,v^n_k$ to $v^{n+1}_h,\,v^{n+1}_k$, respectively, due to an OV relaxation: \newline{} \label{alg:OV}
				$v^{n+1}_h=\xi_ha(V(h(\rho^n_j))-v^{n+1/2}_h), \quad v^{n+1}_k=\xi_ka(V(h(\rho^n_j))-v^n_k)$\;
			}
		}
		Update particle positions according to~\eqref{eq:particle_model}: $x^{n+1}_i=x^n_i+v^{n+1}_i\Delta{t}$, \quad $i=1,\,\dots,\,N$\; \label{alg:update.x}
	}
\end{algorithm}

The implementation of the particle algorithm, which is reported in detail in Algorithm~\ref{alg:MC}, follows closely the stochastic particle model~\eqref{eq:particle_model}. A few comments on some key passages of the algorithm are worthwhile.

The cell size $\Delta{x}$, cf. line~\ref{alg:Dx}, is chosen of the same order of magnitude as the headway function $h$. This way, when considering later interactions among particles within the same cell, cf. line~\ref{alg:pick_particles}, each particle interacts with another particle in a neighbourhood of size comparable to $h$. This is statistically consistent with the Enskog-type kinetic description implemented in the first term on the right-hand side of~\eqref{eq:Enskog}. It is convenient to use a headway function of the form~\eqref{eq:h.bounded} for then we have $h(\rho)=O(c)$ for $\rho=O(1)$. Consequently, with an initial datum like the one depicted in Figure~\ref{fig:initcond} we may easily fulfil the requirement in line~\ref{alg:Dx} of Algorithm~\ref{alg:MC} by setting $\Delta{x}=c$. The value of $c$ should be small enough to further comply with the smallness of $h$ implicitly assumed by the first order approximation~\eqref{eq:approx.f}.

The time step $\Delta{t}$, cf. line~\ref{alg:Dt}, affects both the update of the particle positions in line~\ref{alg:update.x} and the sampling of the random variables $\Theta$ in line~\ref{alg:Bernoulli.Theta} and $\Xi$ in line~\ref{alg:Bernoulli.Xi}. In particular, in the case of slow OV dynamics condition~\eqref{eq:Bernoulli.slow_OV} together with the hyperbolic scaling~\eqref{eq:hyp_scal} produces
\begin{equation}
	\Theta\sim\operatorname{Bernoulli}\!\left(\frac{\Delta{t}}{\epsilon}\right), \qquad
		\Xi\sim\operatorname{Bernoulli}(\Delta{t}),
	\label{eq:Bernoulli_scaling.slow_OV}
\end{equation}
therefore FTL interactions take place at rate $\frac{1}{\epsilon}$, which is much higher than the unitary rate of OV updates and of the hydrodynamic transport in space (line~\ref{alg:update.x}). Conversely, in the case of fast OV dynamics condition~\eqref{eq:Bernoulli.fast_OV} plus the same scaling~\eqref{eq:hyp_scal} produces
\begin{equation}
	\Theta\sim\operatorname{Bernoulli}\!\left(\frac{\Delta{t}}{\epsilon}\right), \qquad
		\Xi\sim\operatorname{Bernoulli}\!\left(\frac{\Delta{t}}{\epsilon}\right),
	\label{eq:Bernoulli_scaling.fast_OV}
\end{equation}
consistently with the fact that now both FTL interactions and OV updates take place at the same much quicker rate than the hydrodynamic transport of the particles (line~\ref{alg:update.x}). In all cases, the condition reported in line~\ref{alg:Dt} is needed in order for all the parameters of the Bernoulli random variables above be actual probabilities. In our subsequent numerical tests we will invariably fix $\Delta{t}=\epsilon$.

Finally, from lines~\ref{alg:switch},~\ref{alg:FTL} we see that interacting particles are ordered in such a way that only the rear vehicle possibly changes speed in consequence of an FTL interaction. This reproduces the ahead-behind anisotropy of vehicle interactions, which causes the coefficient $\frac{1}{2}$ to appear in front of the first term on the right-hand side of the kinetic description~\eqref{eq:Enskog}. Conversely, OV updates are experienced in principle by both interacting particles, cf. line~\ref{alg:OV}, because they are not dictated by pairwise interactions.

\begin{table}[!t]
\centering
\caption{Parameters and functions used in the numerical simulations of Figures~\ref{fig:ARZ},~\ref{fig:LWR}}
\label{tab:param}
\begin{tabular}{c|c|c|c}
\hline
Parameter & Description & Value & Reference equations \\
\hline
\hline
$N$ & total number of particles & $10^6$ & Algorithm~\ref{alg:MC} \\
$c$ & coefficient of the function $h(\rho)$ & $10^{-2}$ & \eqref{eq:h.bounded} \\
$\lambda_0$ & coefficient of the function $\lambda(h)$ & $0.5$ & \eqref{eq:lambda.bounded} \\
$a$ & relaxation parameter & $0.5$ & \eqref{eq:particle_model},~\eqref{eq:inhom_ARZ} \\
$h(\rho)$ & headway function & $\frac{c}{1+\rho}$ & \eqref{eq:h.bounded},~\eqref{eq:p}-\eqref{eq:inhom_ARZ} \\
$\lambda(h)$ & sensitivity of the drivers & $\frac{\lambda_0}{1+h}$ & \eqref{eq:lambda.bounded},~\eqref{eq:p}-\eqref{eq:inhom_ARZ} \\
$V(h)$ & optimal speed function & $\tanh\!\left(\frac{h}{c}\right)$ & \eqref{eq:V.tanh},~\eqref{eq:inhom_ARZ},~\eqref{eq:LWR} \\
$\epsilon$ & scaling parameter & $10^{-1},\,10^{-2},\,10^{-3}$ & Algorithm~\ref{alg:MC} \\
$\Delta{t}$ & time step & $\epsilon$ & Algorithm~\ref{alg:MC} \\
$\Delta{x}$ & cell size & $c$ & Algorithm~\ref{alg:MC} \\
\hline
\end{tabular}
\end{table}

\begin{figure}[!t]
\includegraphics[width=0.33\textwidth]{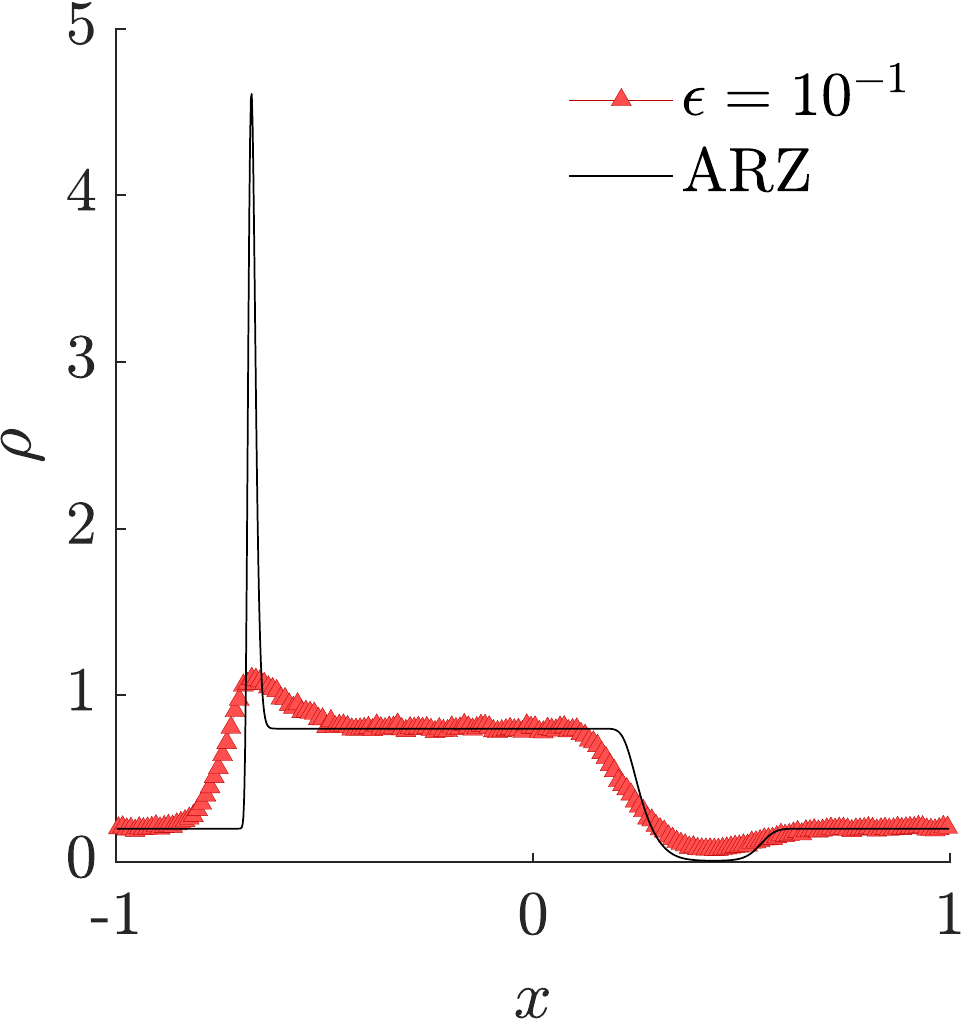}
\includegraphics[width=0.33\textwidth]{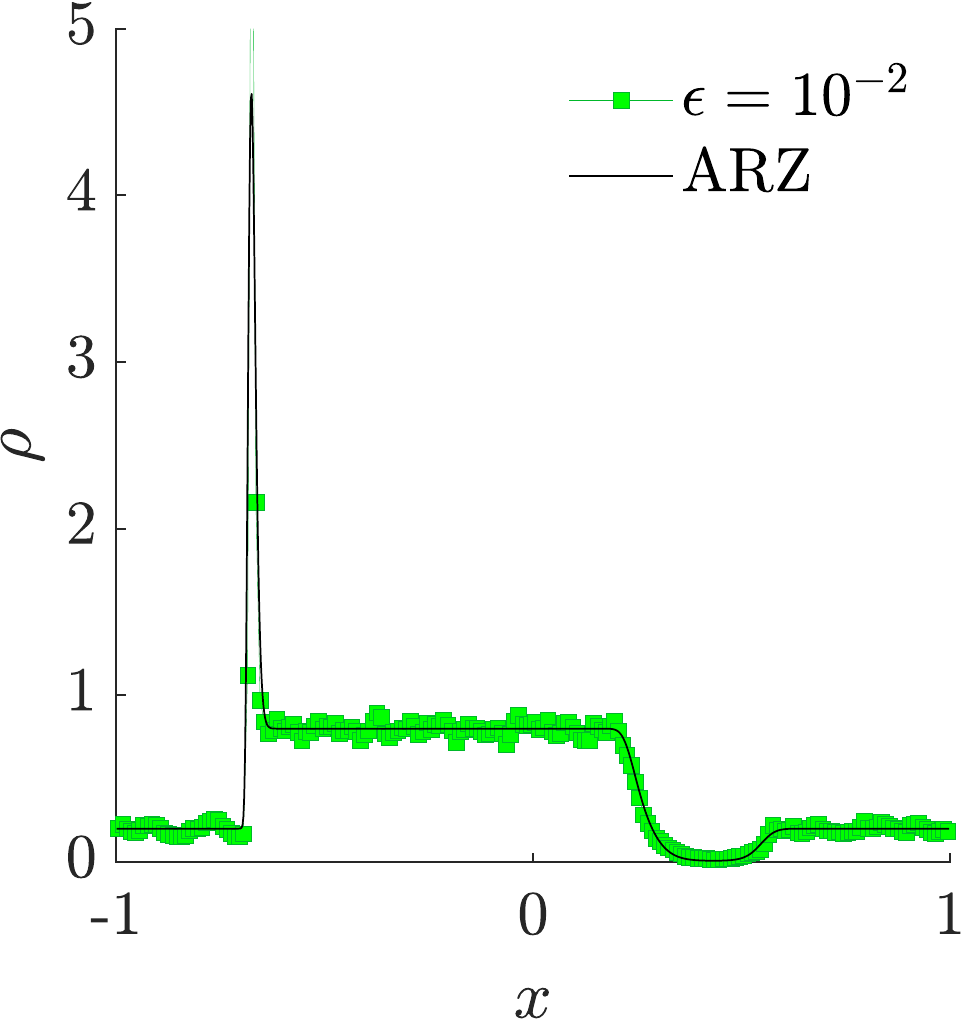}
\includegraphics[width=0.33\textwidth]{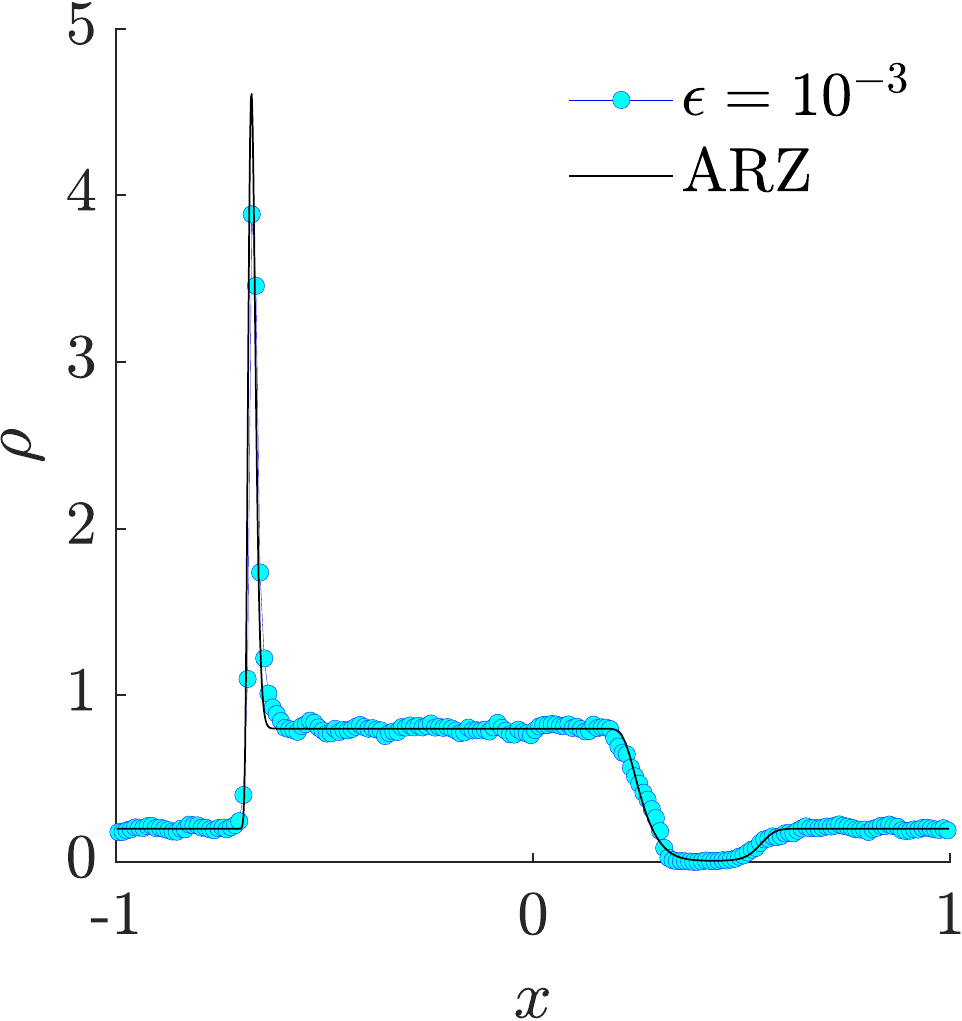} \\
\includegraphics[width=0.33\textwidth]{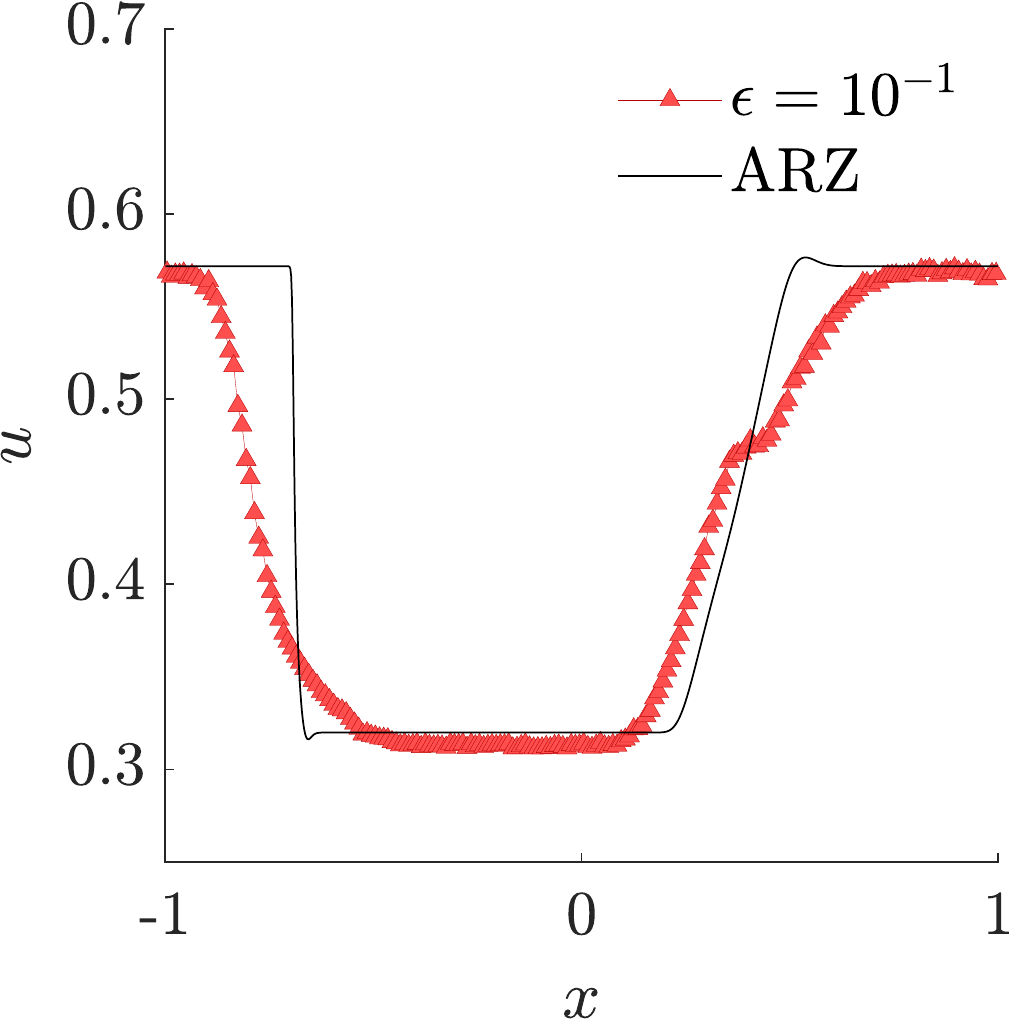}
\includegraphics[width=0.33\textwidth]{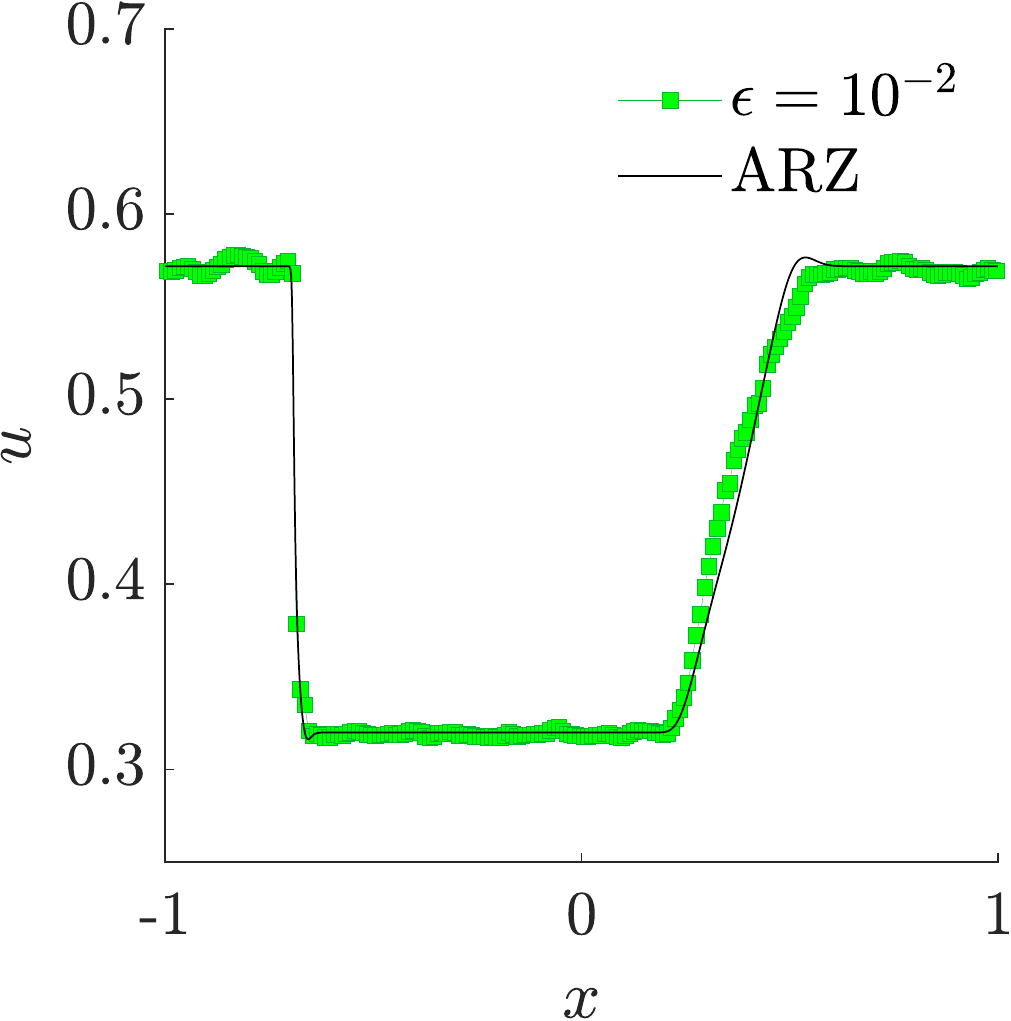}
\includegraphics[width=0.33\textwidth]{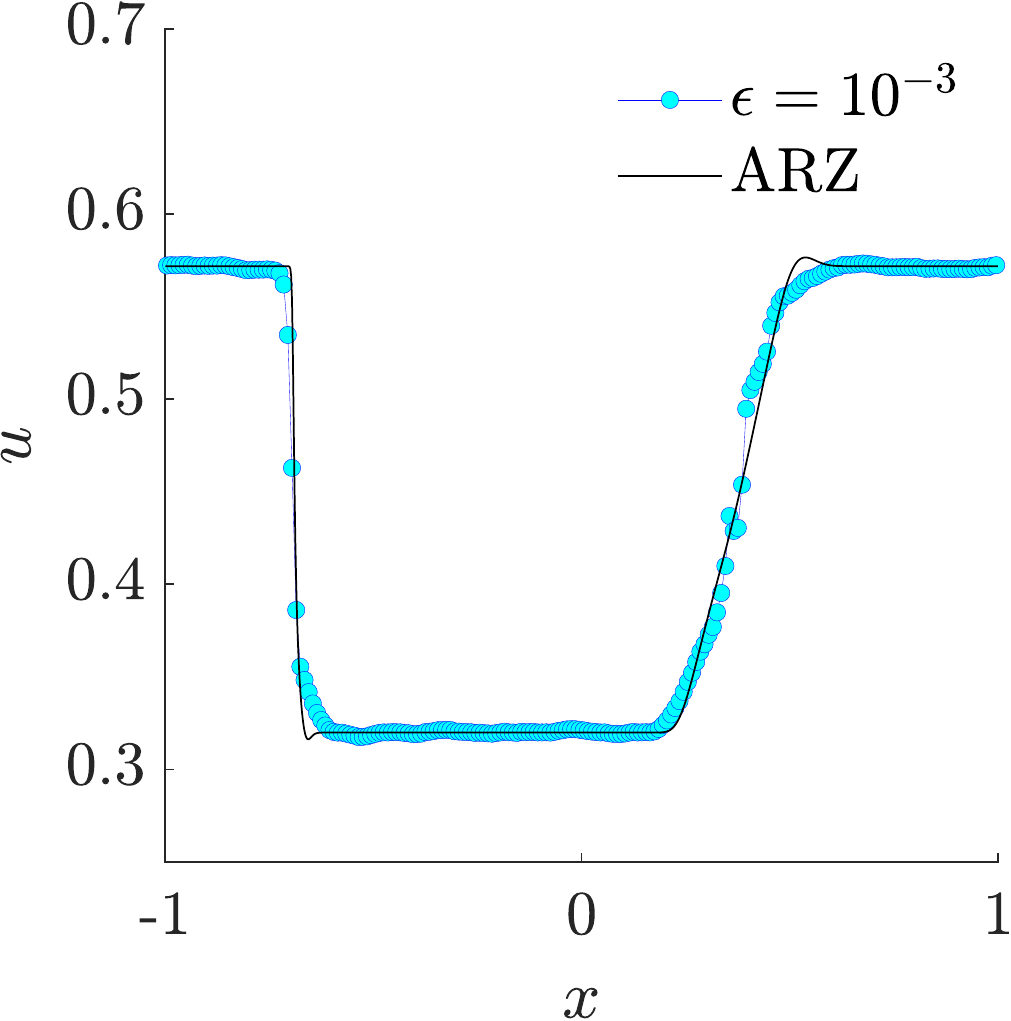}
\caption{Density (top row) and mean speed (bottom row) at time $t=1$ of the inhomogeneous ARZ model (solid line) and of the particle model (markers) in the slow OV regime for $\epsilon$ decreasing from $10^{-1}$ to $10^{-3}$ (left to right)}
\label{fig:ARZ}
\end{figure}

In Figure~\ref{fig:ARZ} we compare the numerical solution at the computational time $t=1$ of the inhomogeneous ARZ model~\eqref{eq:p}-\eqref{eq:inhom_ARZ}, computed by means of a splitting of the transport and relaxation contributions coupled with a classical conservative upwind scheme for the former, with that of the particle model computed by means of Algorithm~\ref{alg:MC} in the regime~\eqref{eq:Bernoulli_scaling.slow_OV}. The initial condition is the one reported in Figure~\ref{fig:initcond}. The other relevant parameters and functions are listed in Table~\ref{tab:param}. It is evident that both the density and the mean speed profiles of the solutions produced by the two models tend to coincide substantially as the scaling parameter $\epsilon$ decreases from $10^{-1}$ to $10^{-3}$. This correctly reproduces the hydrodynamic limit and the related results anticipated by the theory.

\begin{figure}[!t]
\includegraphics[width=0.33\textwidth]{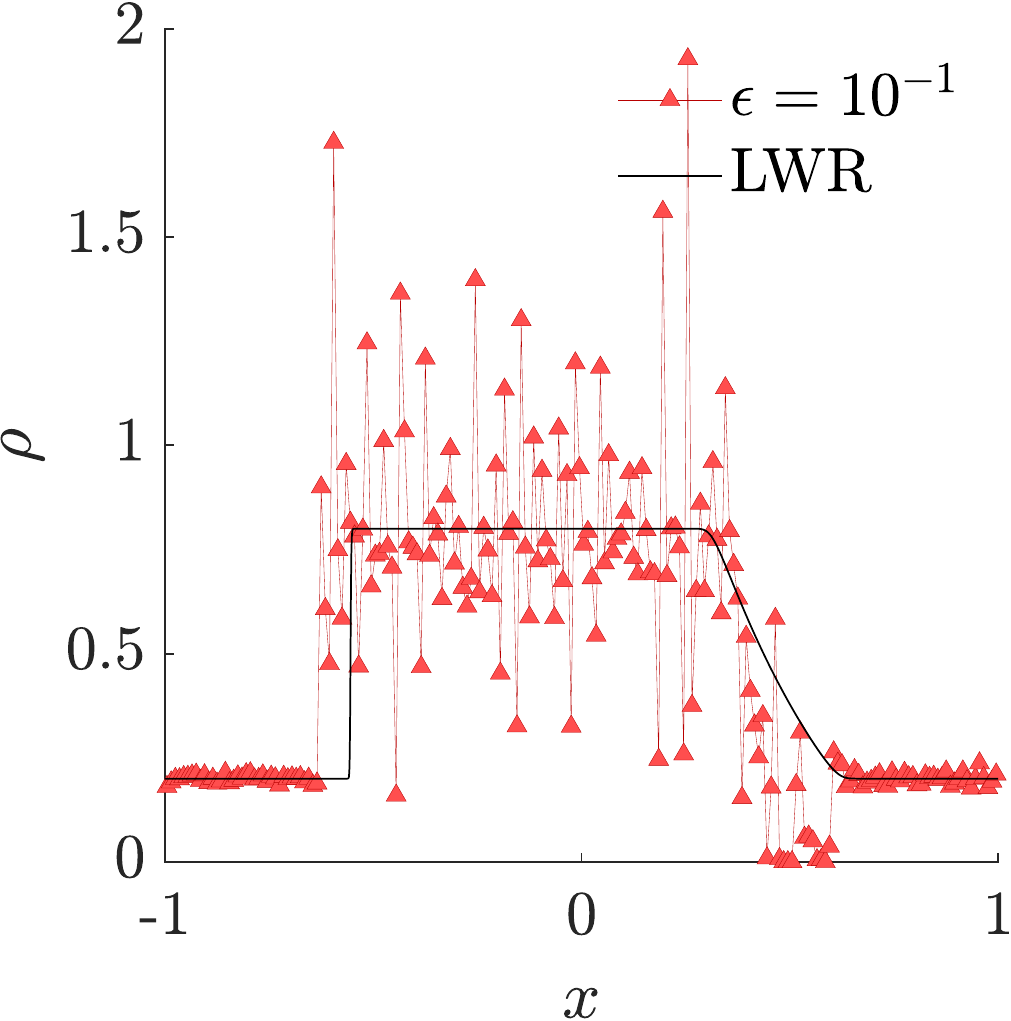}
\includegraphics[width=0.33\textwidth]{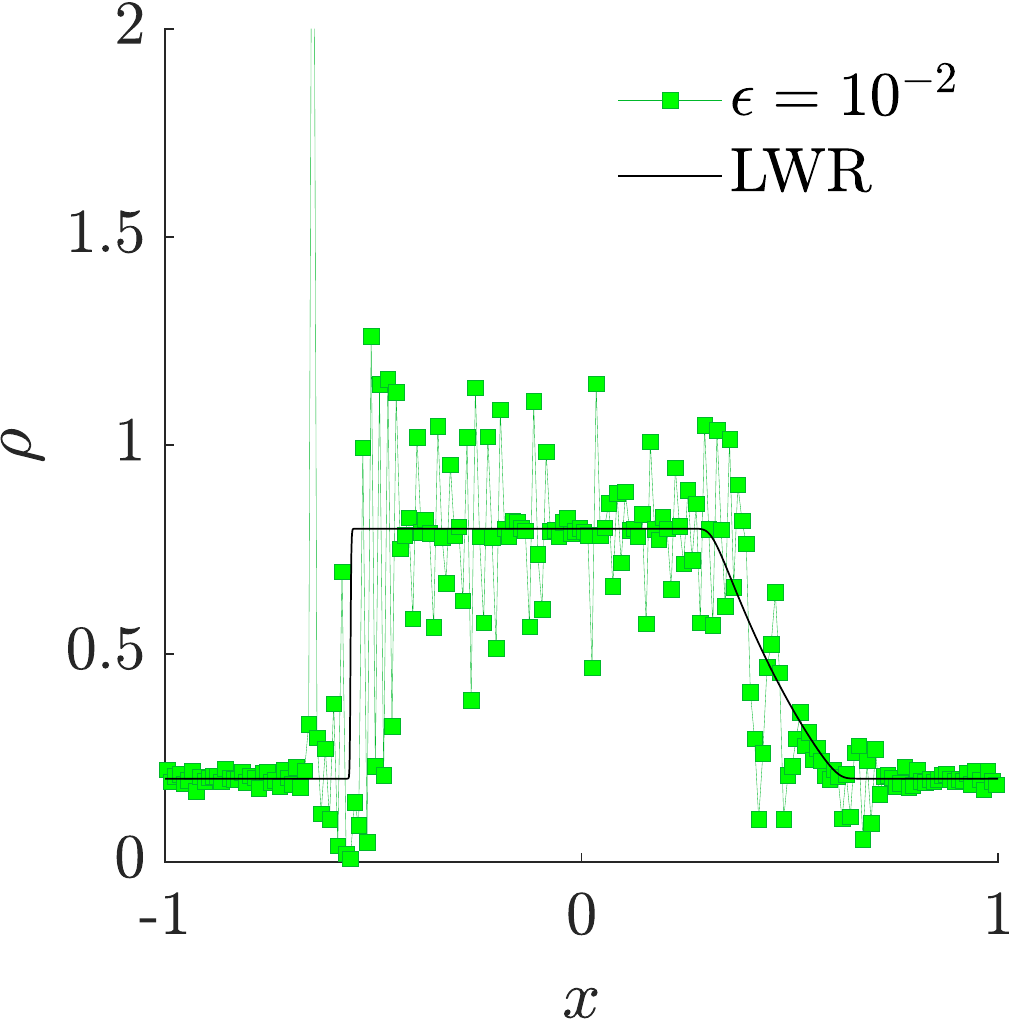}
\includegraphics[width=0.33\textwidth]{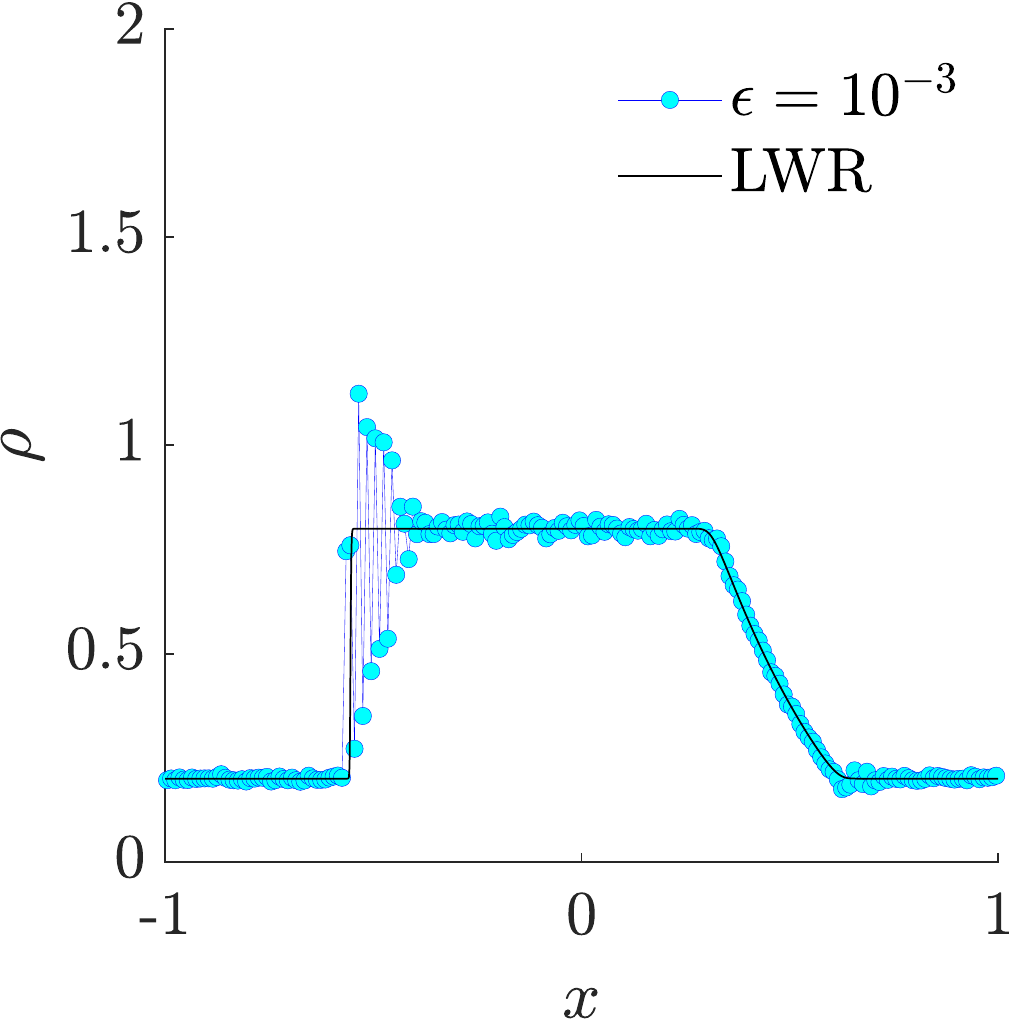} \\
\includegraphics[width=0.33\textwidth]{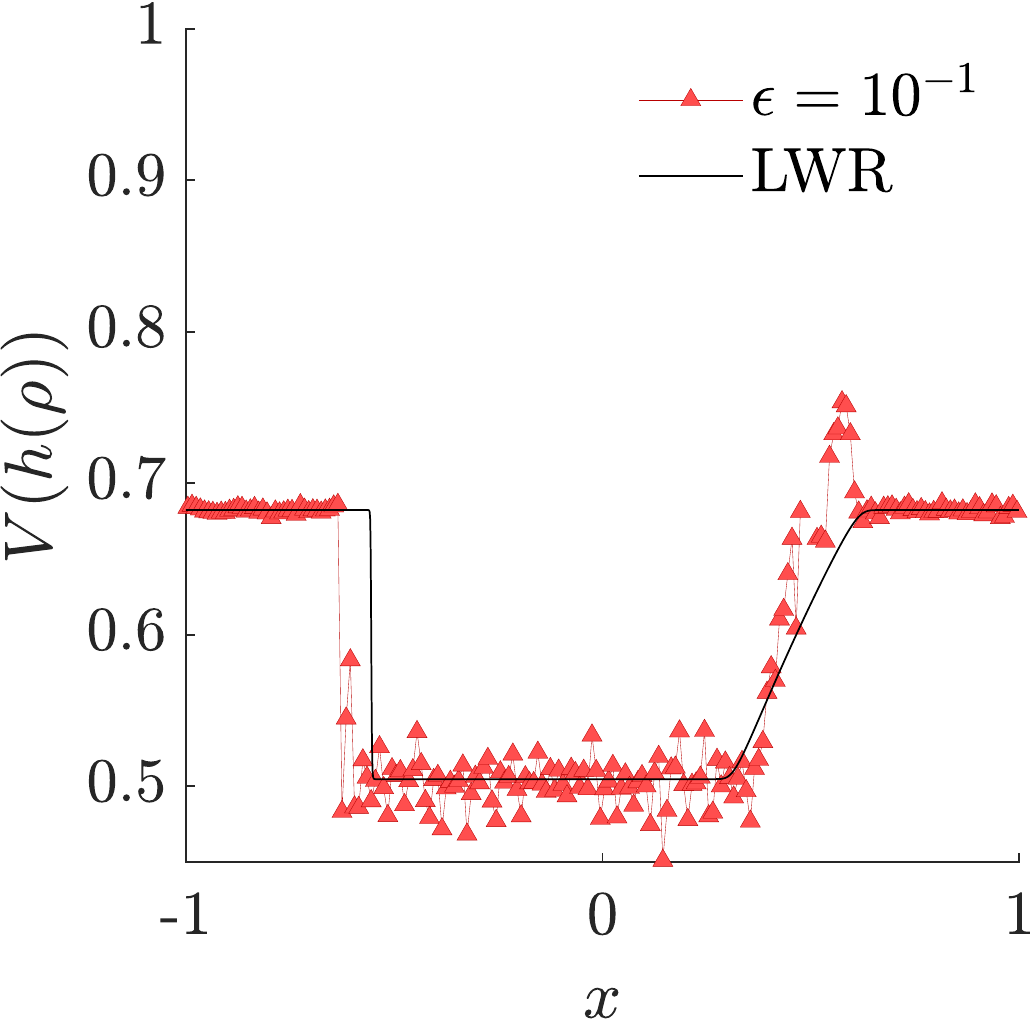}
\includegraphics[width=0.33\textwidth]{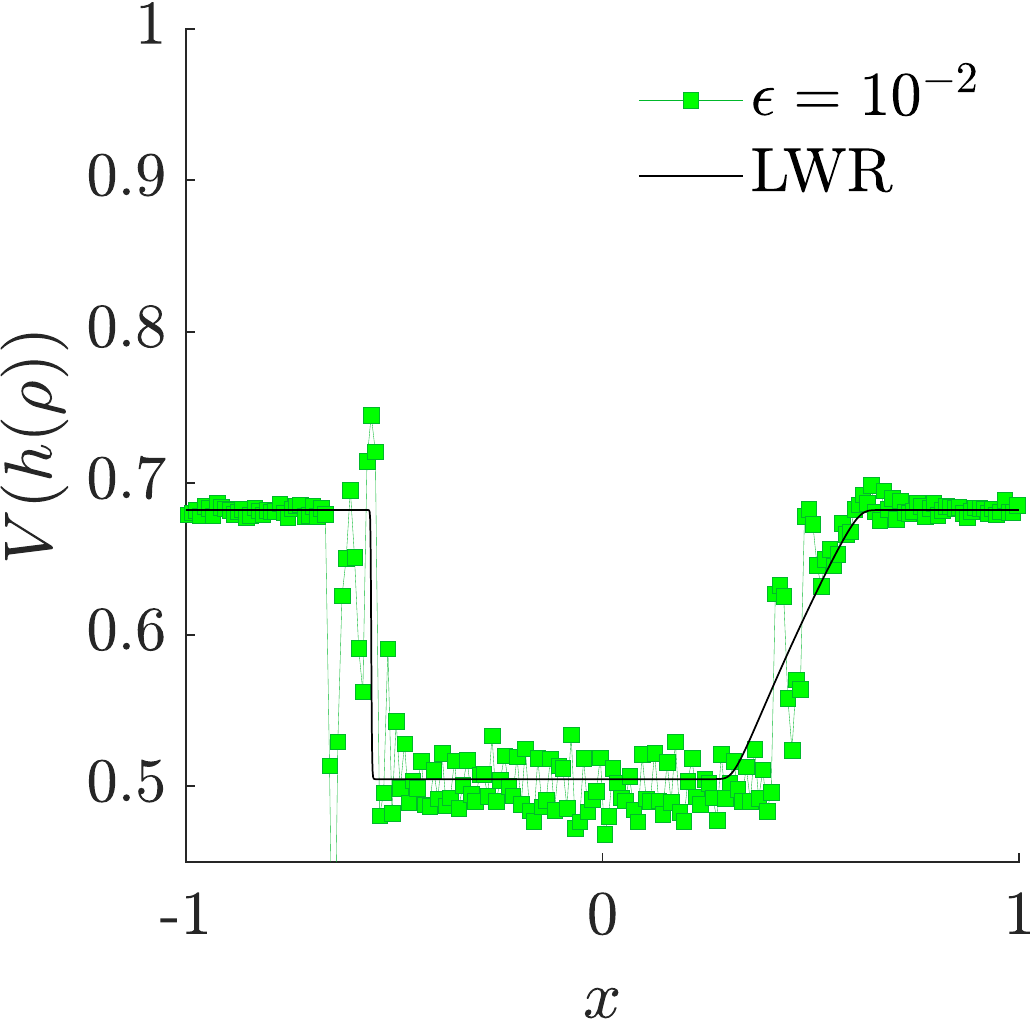}
\includegraphics[width=0.33\textwidth]{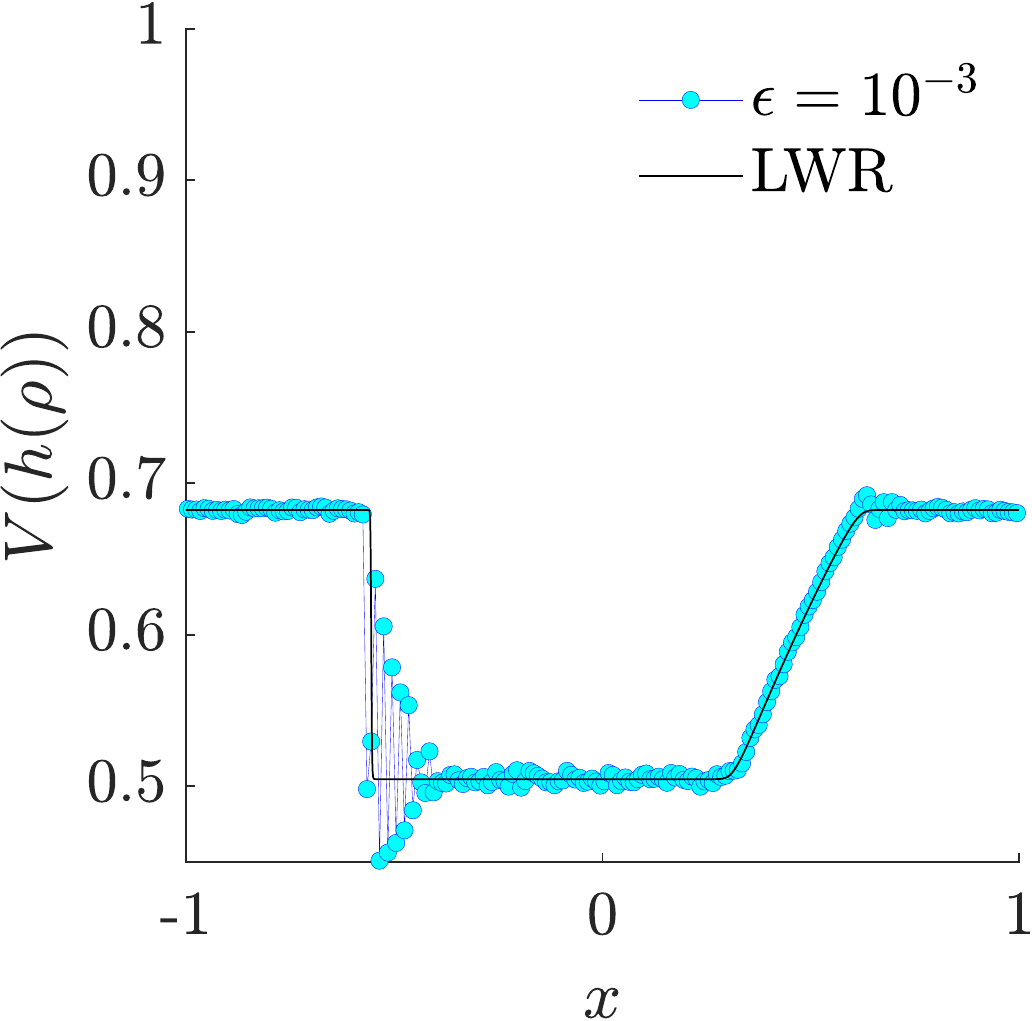}
\caption{Density (top row) and optimal velocity (bottom row) at time $t=1$ of the LWR model (solid line) and of the particle model (markers) in the fast OV regime for $\epsilon$ decreasing from $10^{-1}$ to $10^{-3}$ (left to right)}
\label{fig:LWR}
\end{figure}

In Figure~\ref{fig:LWR} we compare instead the numerical solution at the computational time $t=1$ of the LWR model~\eqref{eq:LWR}, computed by a classical conservative upwind scheme, with that of the particle model computed by means of Algorithm~\ref{alg:MC} in the regime~\eqref{eq:Bernoulli_scaling.fast_OV}, starting again from the initial condition depicted in Figure~\ref{fig:initcond} and using the parameters and functions reported in Table~\ref{tab:param}. In this case, we replace the illustration of the mean speed $u$ with that of the optimal speed $V(h(\rho))$ produced by the local interactions and featured by the LWR scalar conservation law. We observe that the particle solution is now highly oscillatory especially for large $\epsilon$, i.e. far from the true hydrodynamic regime. One should indeed expect a tougher convergence of the \textit{second order} particle dynamics~\eqref{eq:particle_model} to the \textit{first order} macroscopic dynamics~\eqref{eq:LWR}. However, for $\epsilon=10^{-3}$ we notice a nice matching of the two solutions with residual particle oscillations only in correspondence of the rear shock wave, namely where the limit macroscopic solution is not smooth. The frontal rarefaction wave as well as the constant pieces of the solution are instead caught well.

\begin{remark}
Monte Carlo particle methods, such as e.g., the Nanbu-Babovsky~\cite{babovsky1986M2AS,nanbu1980JPSJ} and Bird~\cite{bird1970PF} algorithms, see also~\cite{pareschi2013BOOK}, are usually employed to solve kinetic equations numerically. For instance, in~\cite{dimarco2020JSP} one of such algorithms is used to compute the solution to an Enskog-type equation for vehicular traffic written in an approximate form similar to~\eqref{eq:Enskog.approx}. In our case, instead, Algorithm~\ref{alg:MC} solves the original stochastic particle model~\eqref{eq:particle_model} independently of any theoretical translation into kinetic equations. Therefore, our numerical results of Figures~\ref{fig:ARZ},~\ref{fig:LWR} support genuinely the whole theoretical path from the Enskog-type kinetic description~\eqref{eq:Enskog} of the particle dynamics~\eqref{eq:particle_model} to the approximate Enskog-type description~\eqref{eq:Enskog.approx} and finally to the hydrodynamic limits~\eqref{eq:inhom_ARZ},~\eqref{eq:LWR} in the respective slow and fast OV regimes.
\end{remark}

\subsection{Stability of the inhomogeneous ARZ model}
We test now numerically the macroscopic stability condition~\eqref{eq:stability} of the inhomogeneous ARZ model~\eqref{eq:p}-\eqref{eq:inhom_ARZ}. We consider the same functions and parameters listed in Table~\ref{tab:param}, for which condition~\eqref{eq:stability} becomes explicitly
$$ \operatorname{sech}^2\!\left(\frac{1}{1+\rho_0}\right)\leq \frac{c\lambda_0}{2}\cdot\frac{(1+\rho_0)^2}{1+\rho_0+c}, $$
$\rho_0\in [0,\,1]$ being the constant density of uniform traffic whose stability is being studied. Choosing $\rho_0=0.5$ and letting $c=10^{-2}$, $\lambda_0=0.5$ like in Table~\ref{tab:param} yields
$$ \operatorname{sech}^2\!\left(\frac{2}{3}\right)\leq\frac{9}{2416}, $$
which is violated because $\operatorname{sech}^2\!\left(\frac{2}{3}\right)\approx 0.66$ while $\frac{9}{2416}\approx 0.004$. We expect therefore a progressive growth in time of any initial perturbation to the constant state $(\rho_0,\,V(h_0))=(0.5,\,\tanh\!\left(\frac{2}{3}\right))$.

\begin{figure}[!t]
\centering
\includegraphics[width=0.33\textwidth]{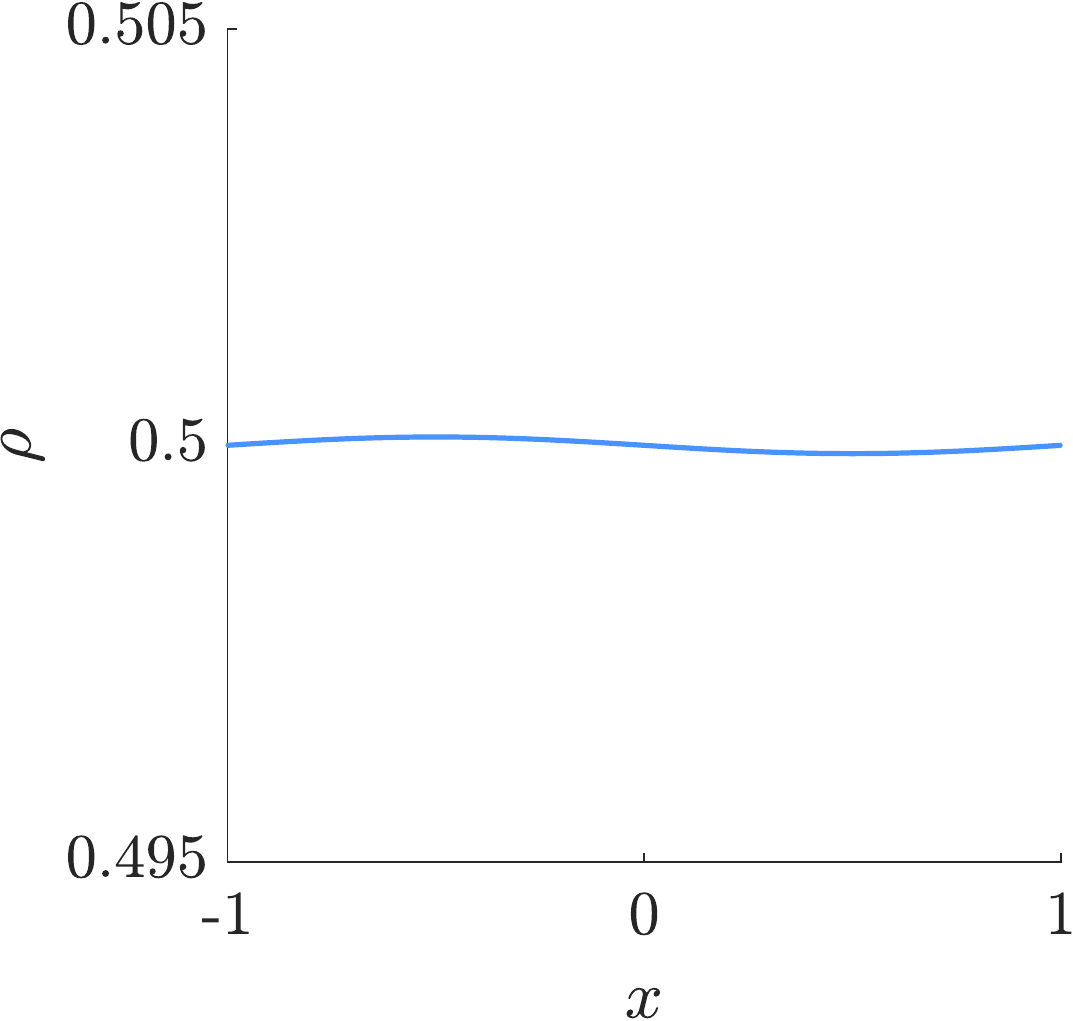}
\includegraphics[width=0.33\textwidth]{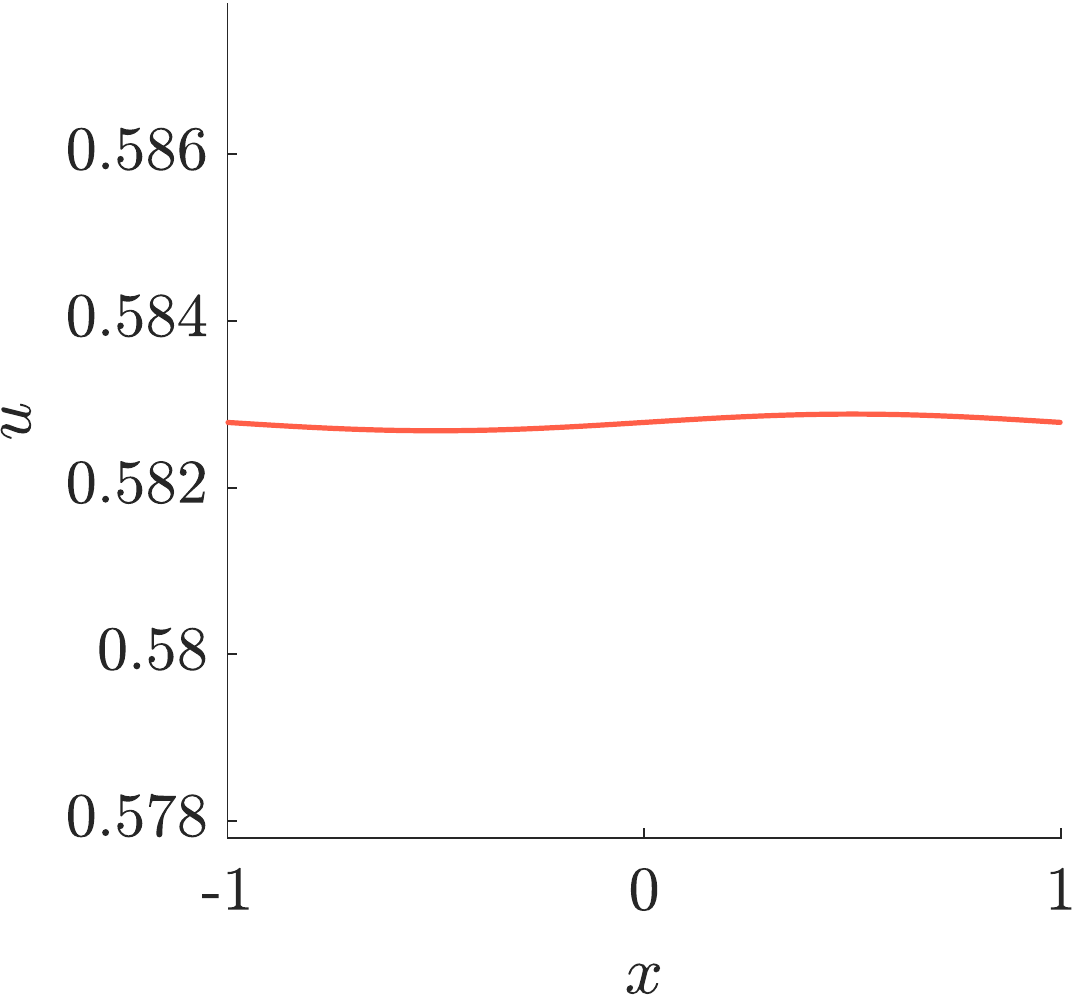}
\caption{Initial condition for the inhomogeneous ARZ model given as a small perturbation of the constant state $\rho_0=0.5$, $u_0=V(h(\rho_0))=\tanh\!\left(\frac{1}{1+\rho_0}\right)=\tanh\!\left(\frac{2}{3}\right)\approx 0.583$}
\label{fig:stability-initcond}
\end{figure}
\begin{figure}[!t]
\centering
\begin{minipage}[b]{\textwidth}
	\includegraphics[width=0.32\textwidth]{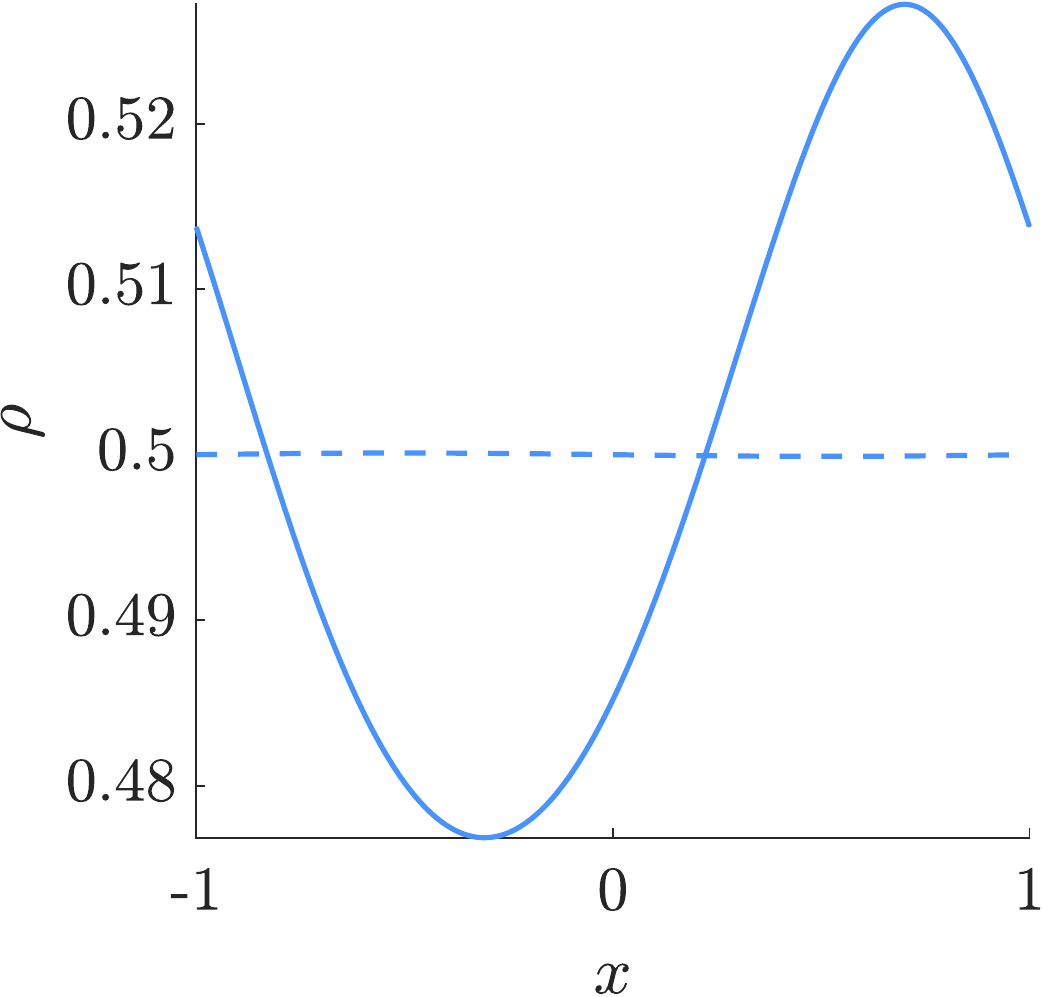}
	\includegraphics[width=0.32\textwidth]{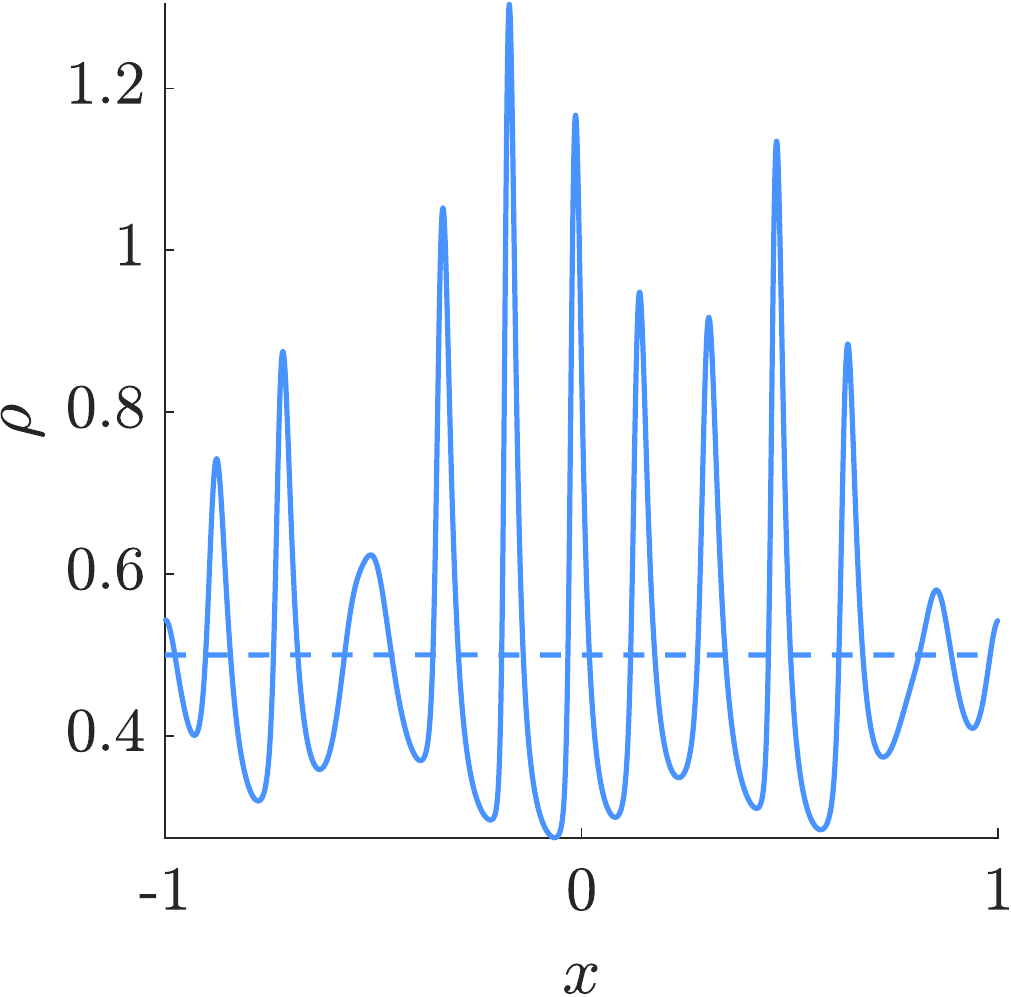}
	\includegraphics[width=0.32\textwidth]{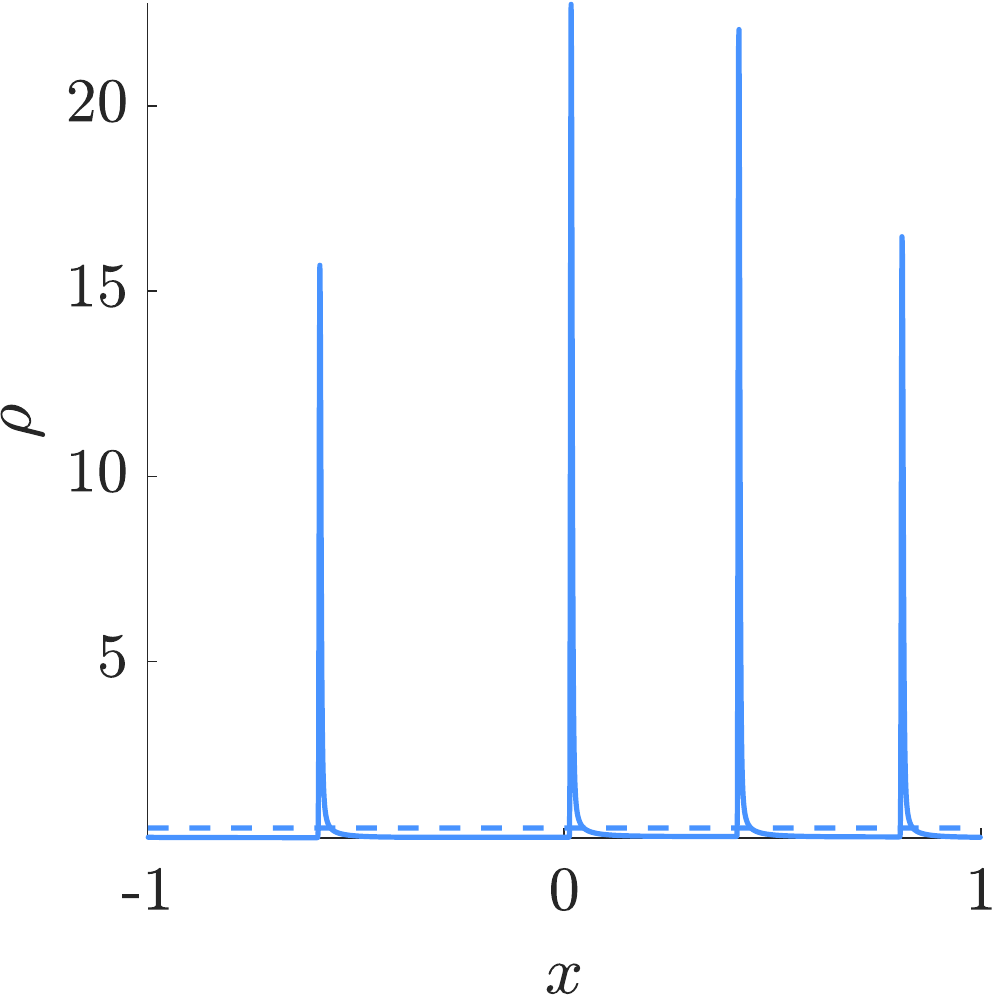}
\end{minipage}
\begin{minipage}[b]{\textwidth}
	\subfigure[$a=0.1$]{\includegraphics[width=0.32\textwidth]{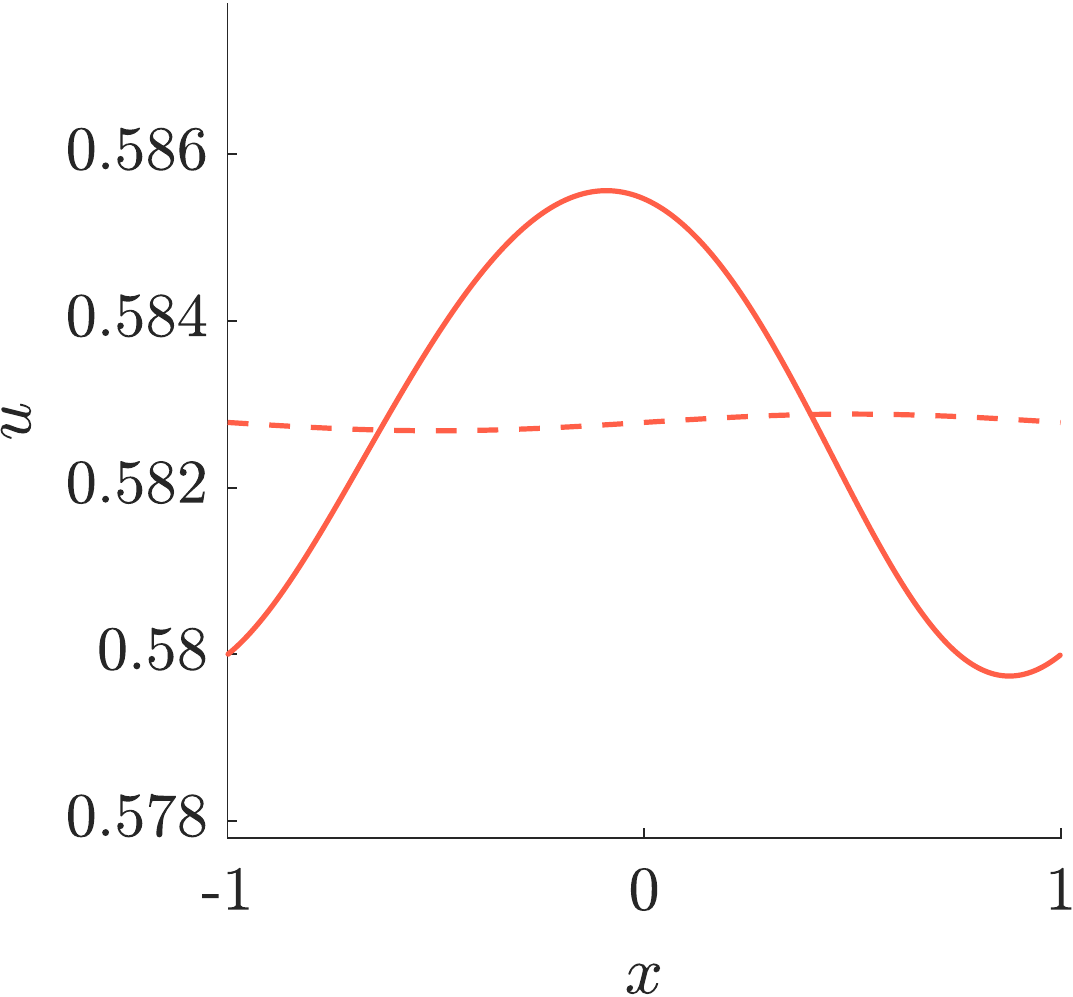}}
	\subfigure[$a=1$]{\includegraphics[width=0.32\textwidth]{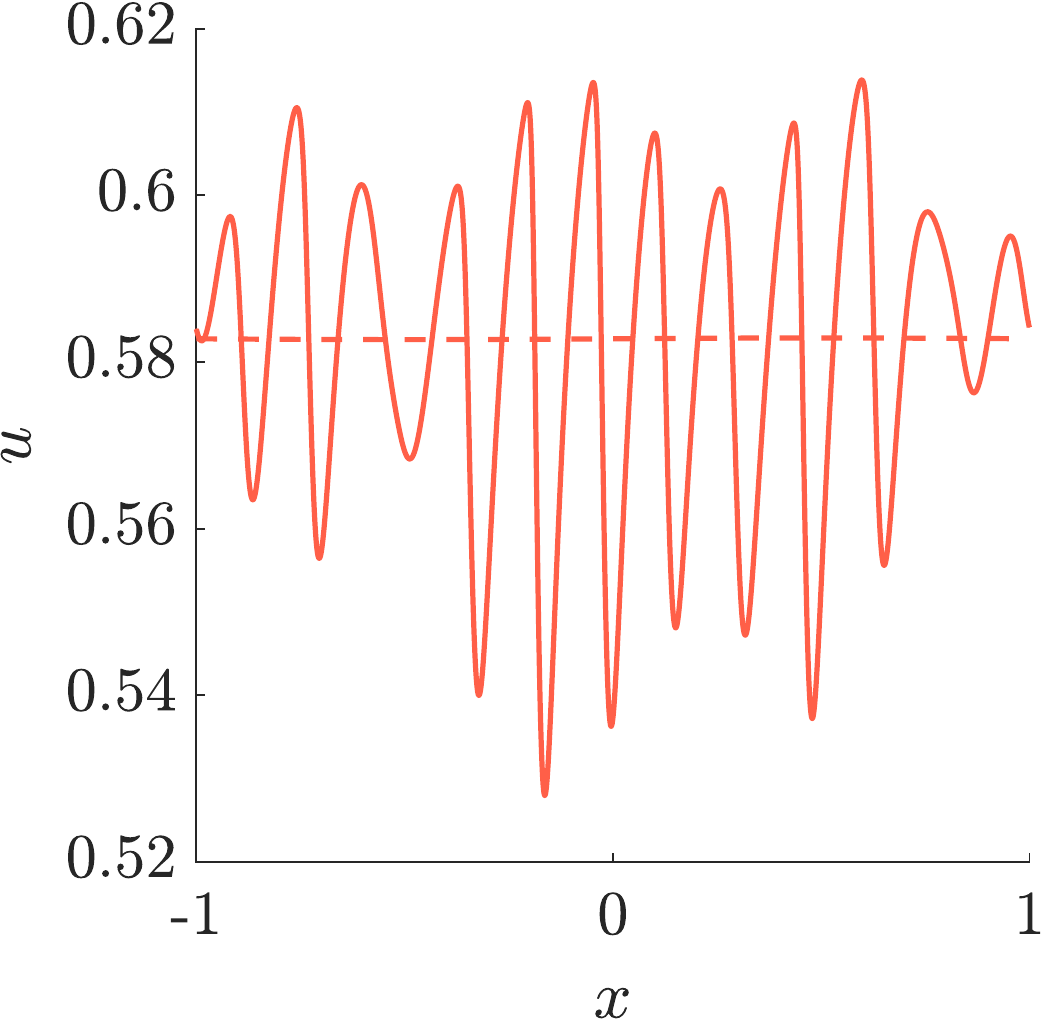}}
	\subfigure[$a=10$]{\includegraphics[width=0.32\textwidth]{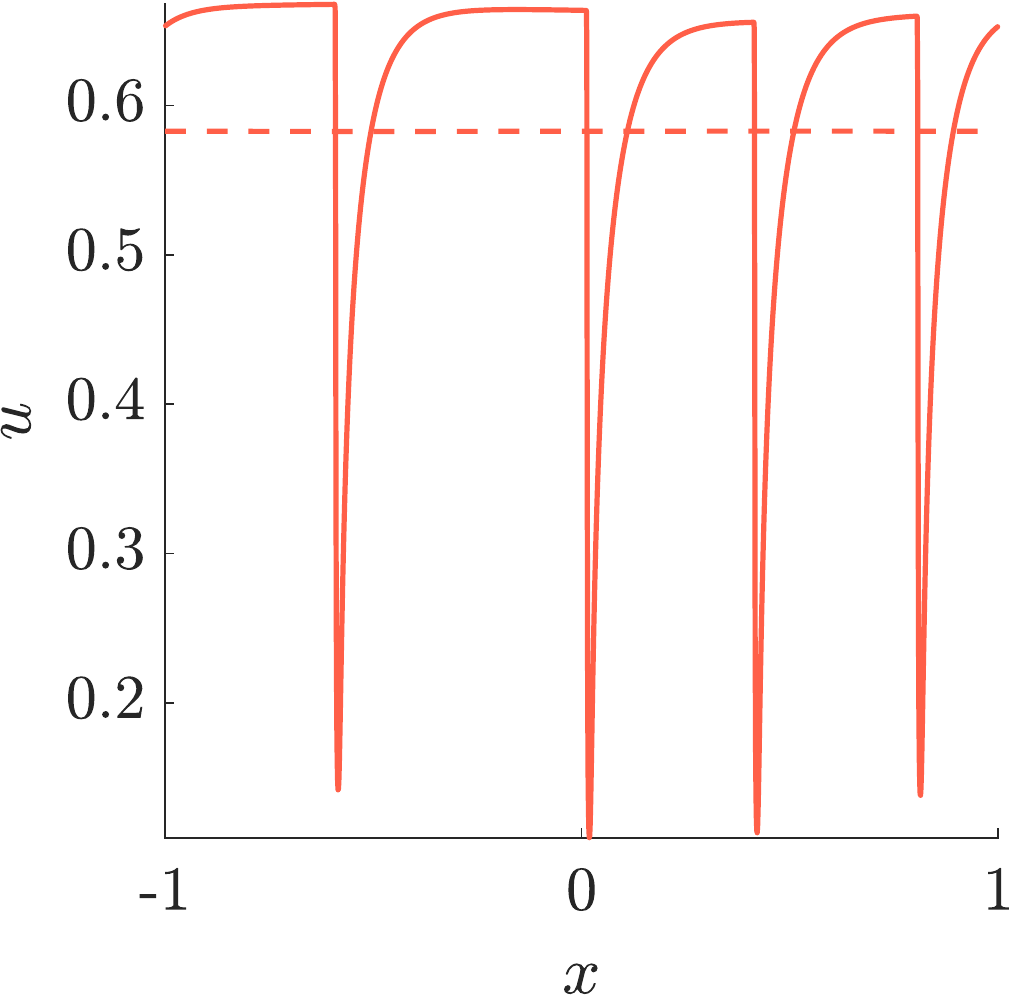}}
\end{minipage}
\caption{Unstable regime of the inhomogeneous ARZ model. Solid lines: amplification of the initial perturbation at time $t=40$ for increasing values of the relaxation rate $a$. Dashed lines: initial condition (cf. Figure~\ref{fig:stability-initcond}) reported for comparison}
\label{fig:stability-unstable}
\end{figure}

For our numerical test we consider, in particular, wave profiles of the form~\eqref{eq:perturbations} with $\eta=10^{-4}$ and initially
$$ \hat{\rho}(x,0)=-\sin{(\pi x)}, \qquad \hat{u}(x,0)=\sin{(\pi x)}, $$
which represent perturbations with opposite phases mimicking a lower mean speed where the density is higher and vice versa, see Figure~\ref{fig:stability-initcond}. The evolution at the computational time $t=40$ is depicted in Figure~\ref{fig:stability-unstable} for increasing values of the relaxation rate from $a=0.1$ to $a=1$ and $a=10$. As predicted by the theory of Section~\ref{sect:stability}, this parameter does not affect the stability of the uniform traffic flow and indeed Figure~\ref{fig:stability-unstable} shows an amplification of the initial perturbation in all the considered cases. However, the value of $a$ affects considerably the waveform into which the initial perturbation evolves. In particular, for a sufficiently large value of $a$ we observe the appearance of \textit{jamitons}, namely travelling waves in the form of localised peaks of high traffic density within a globally moderate mean density along the road, cf. Figure~\ref{fig:stability-unstable}c. This type of instability of the inhomogeneous ARZ model was already observed in~\cite{ramadan2021SEMAI-SIMAI}; here we link its onset to a specific parameter of the microscopic interactions among the vehicles.

\begin{figure}[!t]
\centering
\begin{minipage}[b]{\textwidth}
	\centering
	\includegraphics[width=0.32\textwidth]{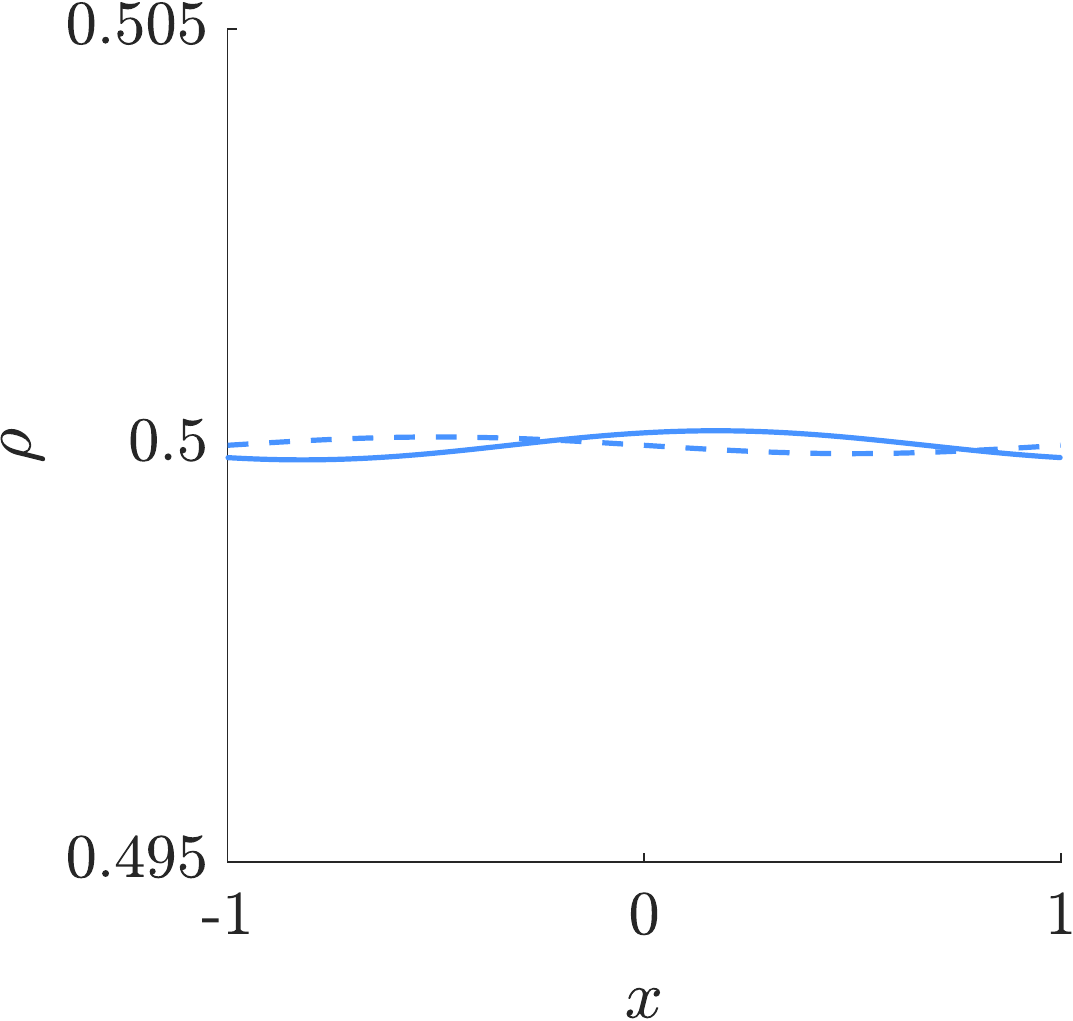}
	\includegraphics[width=0.32\textwidth]{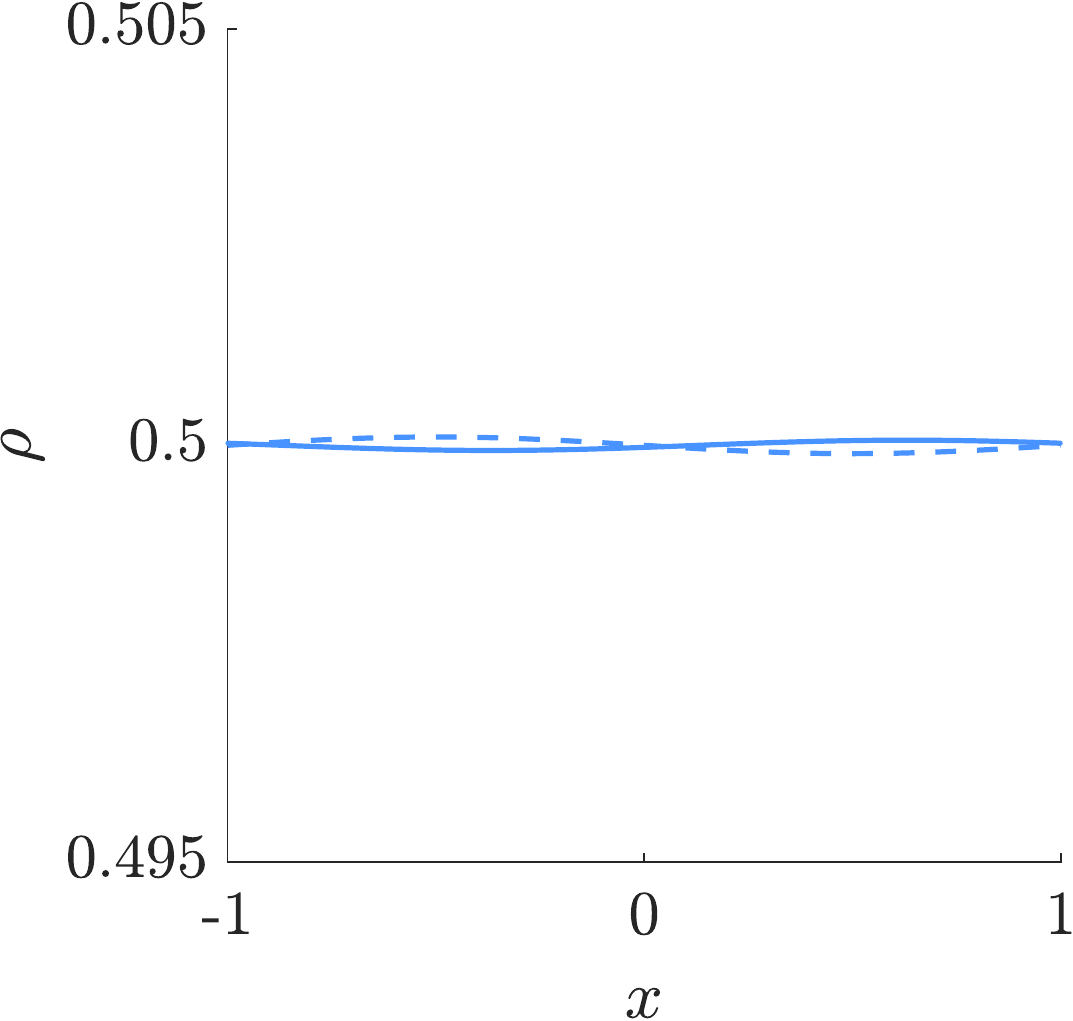}
	\includegraphics[width=0.32\textwidth]{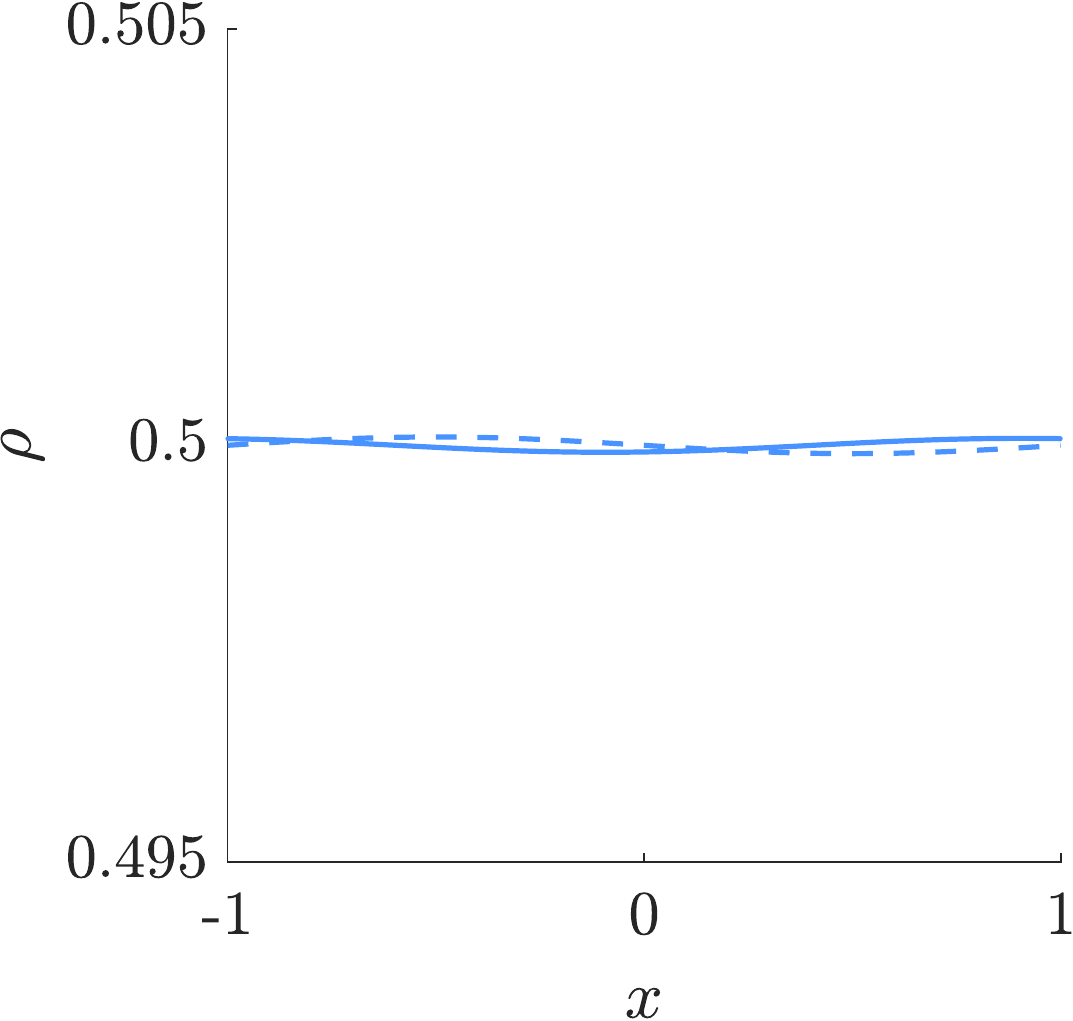}
\end{minipage}
\begin{minipage}[b]{\textwidth}
	\centering
	\subfigure[$a=0.1$]{\includegraphics[width=0.32\textwidth]{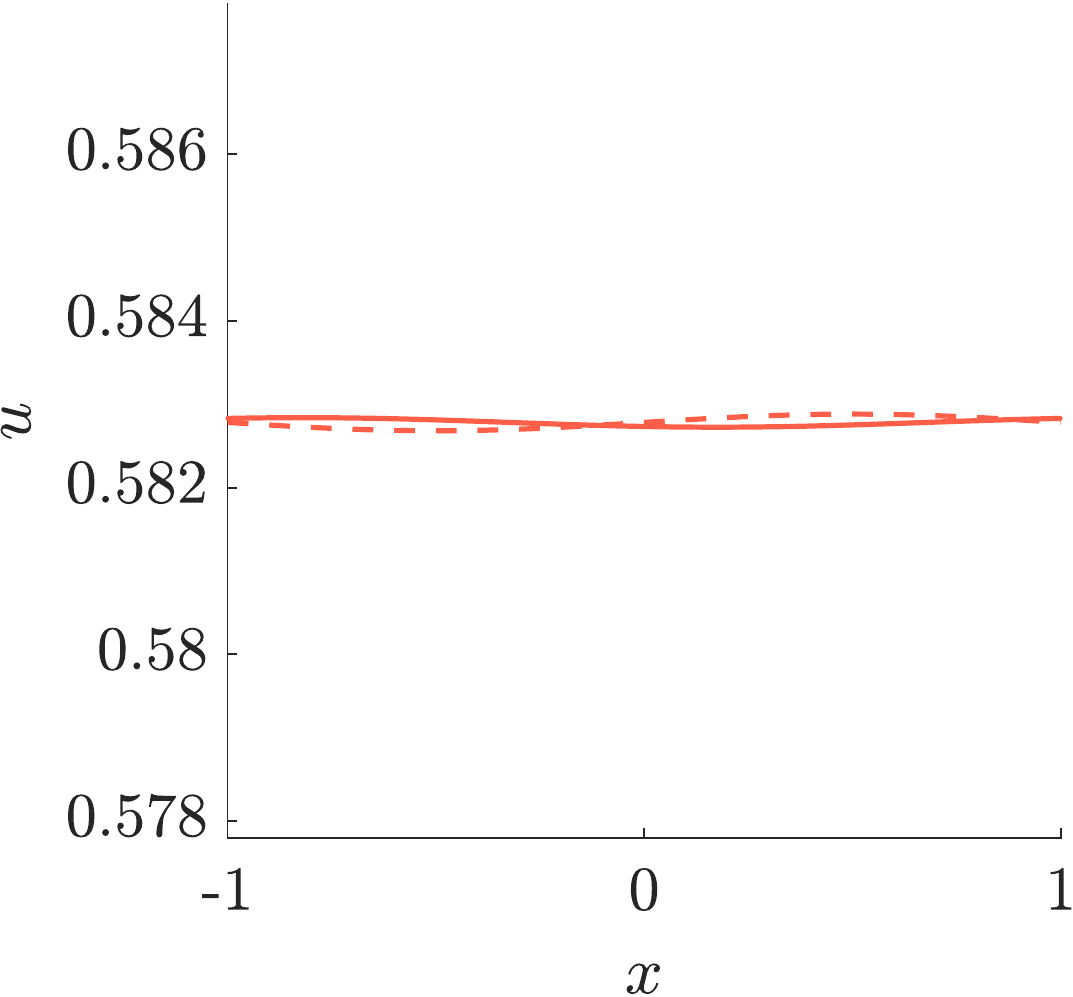}}
	\subfigure[$a=1$]{\includegraphics[width=0.32\textwidth]{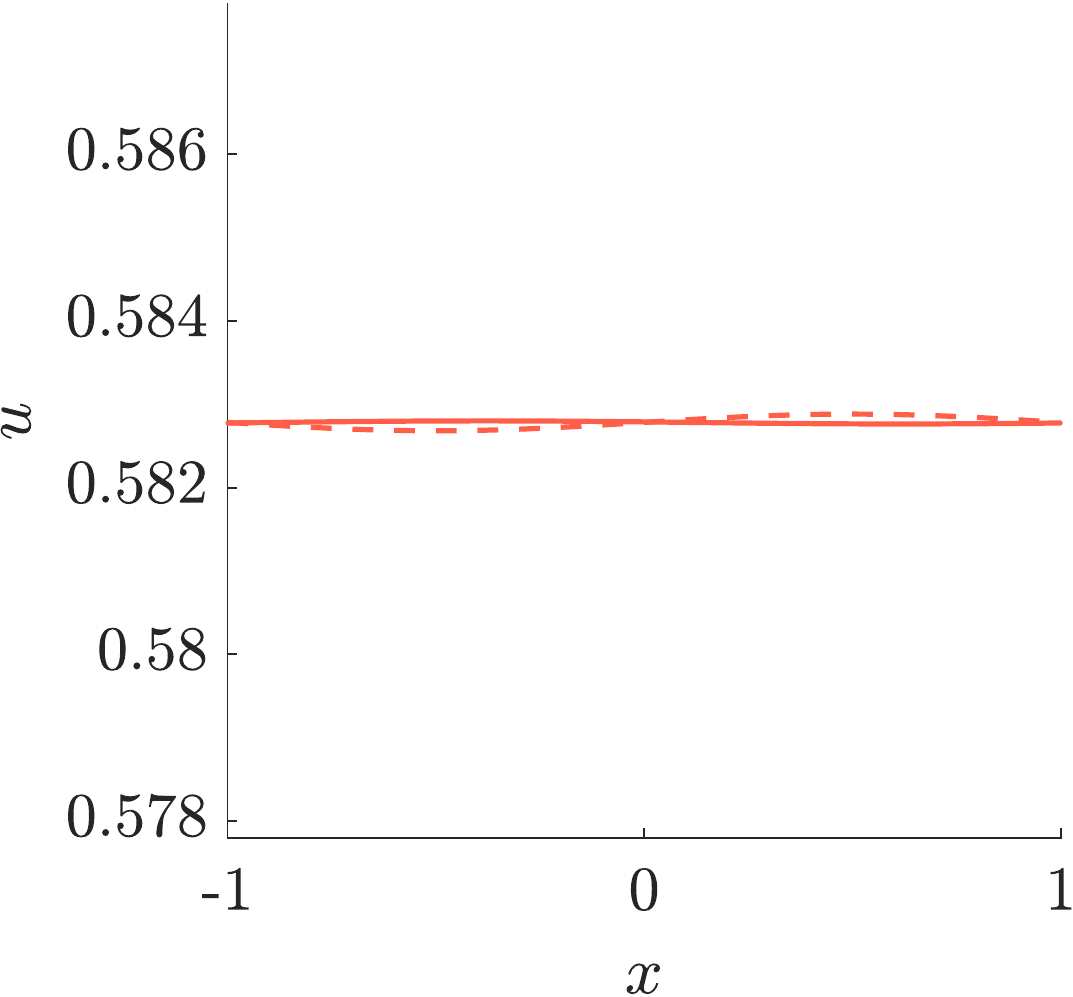}}
	\subfigure[$a=10$]{\includegraphics[width=0.32\textwidth]{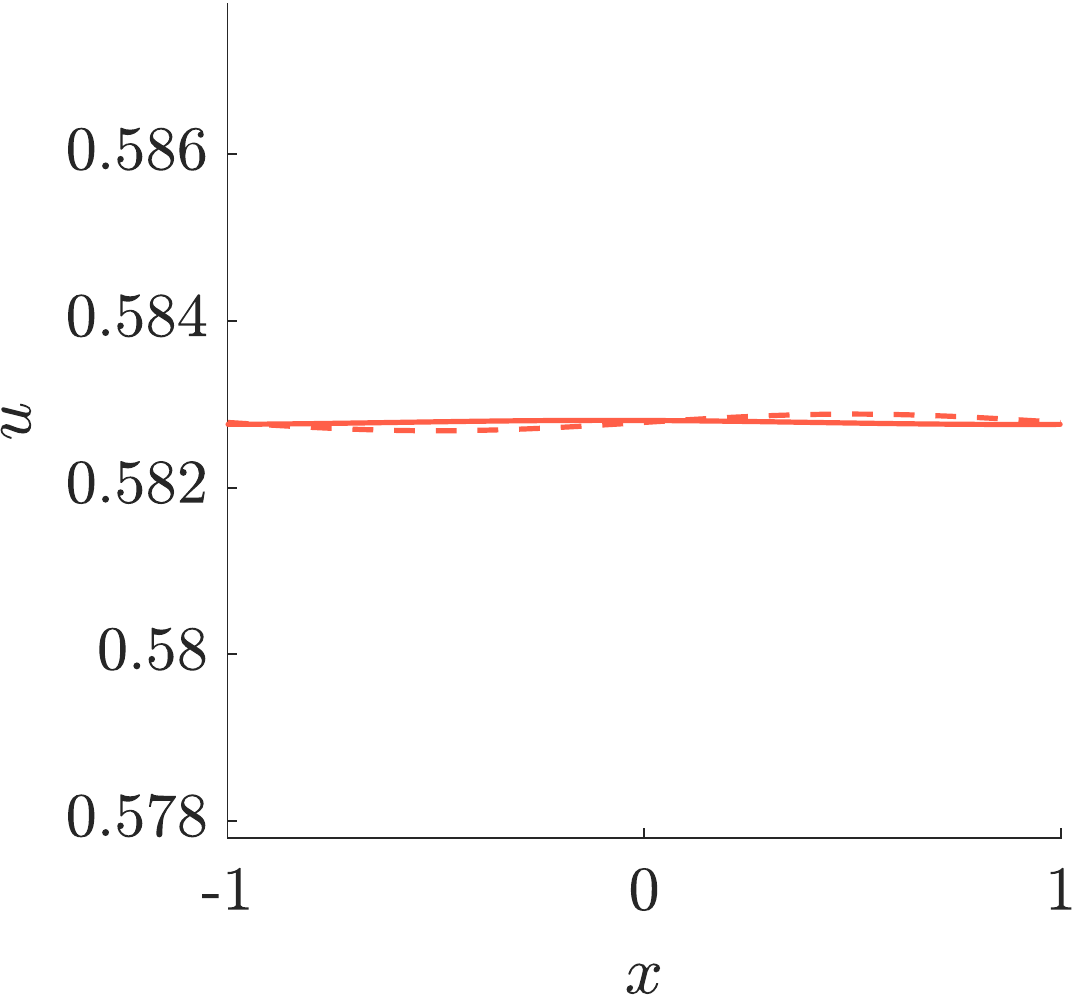}}
\end{minipage}
\caption{Stable regime of the inhomogeneous ARZ model. Solid lines: evolution of the initial perturbation at time $t=40$ for increasing values of the relaxation rate $a$. Dashed lines: initial condition (cf. Figure~\ref{fig:stability-initcond}) reported for comparison}
\label{fig:stability-stable}
\end{figure}

By increasing the parameter $\lambda_0$ to $\lambda_0=100$, with all other parameters and functions unchanged, the stability condition~\eqref{eq:stability} becomes
$$ \operatorname{sech}^2\!\left(\frac{2}{3}\right)\leq\frac{275}{302}, $$
which is now satisfied because $\frac{275}{302}\approx 0.91$. In this case, we observe in Figure~\ref{fig:stability-stable} a stable evolution of the initial perturbation at the computational time $t=40$ independently of the relaxation rate $a$, as predicted by the theory.

\section{Conclusions}
\label{sec:conclusions}
Statistical mechanics and kinetic theory provide an effective theoretical paradigm on which to ground the derivation of macroscopic traffic models out of car-following vehicle dynamics. In this paper, we have shown that they allow one to ascertain the microscopic origins of different macroscopic descriptions of traffic trends in terms of fundamental features of the elementary processes involved in pairwise vehicle interactions. In particular, we have found that the very same microscopic car-following model, consisting of the superposition of FTL interactions and OV relaxation, produces aggregate trends which are well described by either a second order inhomogeneous ARZ model or a first order LWR model depending on the order of magnitude of the relative rate of FTL and OV dynamics. Moreover, we have shown that the aggregate stability of a uniform traffic flow depends on the driver sensitivity to FTL dynamics but not on the OV relaxation rate. The latter affects however considerably the shape of the density waves in the unstable regime, possibly leading to the formation of jamitons.

In our view, compared to other micro-to-macro upscaling procedures proposed in the literature our approach has the merit of elucidating effectively and bridging rigorously the fundamental physical features of traffic flow responsible for certain aggregate trends and the analytical results in terms of macroscopic traffic equations.

\section*{Acknowledgements}
The research of B.P. is based upon work supported by the U.S. Department of Energy's Office of Energy Efficiency and Renewable Energy (EERE) under the Vehicle Technologies Office award number CID DE-EE0008872. The views expressed herein do not necessarily represent the views of the U.S. Department of Energy or the United States Government.

B.P.'s work was partially supported by the National Science Foundation under Cyber-Physical Systems Synergy Grant No. CNS-1837481.

A.T.'s work was partially supported  by the Italian Ministry for Education, University and Research (MIUR) through the ``Dipartimenti di Eccellenza'' Programme (2018-2022), Department of Mathematical Sciences ``G. L. Lagrange'', Politecnico di Torino (CUP: E11G18000350001) and through the PRIN 2017 project (No. 2017KKJP4X) ``Innovative numerical methods for evolutionary partial differential equations and applications''.
	
	
F.A.C. and A.T. are members of GNFM (Gruppo Nazionale per la Fisica Matematica) of INdAM (Istituto Nazionale di Alta Matematica), Italy.
	
\bibliographystyle{plain}
\bibliography{CfPbTa-FTL+OV}
\end{document}